\documentclass[aps,prd,reprint,amsmath,amssymb,superscriptaddress,nofootinbib,floatfix]{revtex4-2}
\usepackage{bm}
\usepackage{graphicx}
\usepackage{xcolor}
\usepackage{hyperref}
\usepackage{booktabs}
\usepackage{multirow}
\usepackage{tikz}
\usepackage{comment} 
\usepackage[compat=1.1.0]{tikz-feynman}
\usetikzlibrary{decorations.markings,decorations.pathmorphing,calc}

\newcommand{\keV}{\,\mathrm{keV}}
\newcommand{\MeV}{\,\mathrm{MeV}}
\newcommand{\GeV}{\,\mathrm{GeV}}
\newcommand{\Db}{\bar{D}}
\newcommand{\Dbs}{\bar{D}^{*}}
\newcommand{\Sc}{\Sigma_c}
\newcommand{\Scs}{\Sigma_c^{*}}
\newcommand{\Lc}{\Lambda_c}
\newcommand{\Pc}{P_c}
\newcommand{\Pcp}{P_c'}
\newcommand{\eps}{\epsilon}
\newcommand{\bK}{\bm{K}}
\newcommand{\beps}{\bm{\epsilon}}
\newcommand{\bS}{\bm{S}}
\newcommand{\half}{\tfrac12}

\tikzfeynmanset{
  vector/.style={gluon},
  pseudoscalar/.style={scalar},
  baryon/.style={fermion},
}
\tikzset{
  dblplain/.style={double,double distance=1.25pt,line width=0.3pt,
     postaction={decorate},
     decoration={markings,mark=at position 0.56 with {\arrow{Stealth}}}},
  dblspect/.style={double,double distance=1.25pt,line width=0.3pt},
  plainspect/.style={postaction={decorate},
     decoration={markings,mark=at position 0.6 with {\arrow{Stealth}}}},
  phot/.style={decorate,decoration={snake,amplitude=1.5pt,segment length=4.8pt},
     line width=0.5pt},
}
\newcommand{\pcf}[1]{%
  \coordinate (L)  at (0,0);   \coordinate (TR) at (1.95,1.12);
  \coordinate (BR) at (1.95,-1.12); \coordinate (G)  at (3.35,1.12);
  \coordinate (F)  at (3.35,-1.12); \coordinate (IN) at (-1.5,0);
  \draw[dblplain] (IN) -- (L);  \draw[plainspect] (BR) -- (F);  \draw[phot] (TR) -- (G);
  \node[left,font=\scriptsize]  at (IN) {\(P_c'\)};
  \node[right,font=\scriptsize] at (F)  {\(P_c\)};
  \node[above right=-1pt,font=\scriptsize] at (G) {\(\gamma\)};
  \node[font=\footnotesize] at (1.0,-1.85) {(#1)};}
\newcommand{\lUL}[1]{\node[font=\scriptsize] at ($(L)!0.5!(TR)+(-0.27,0.17)$) {#1};}
\newcommand{\lV}[1] {\node[font=\scriptsize] at ($(TR)!0.5!(BR)+(0.32,0)$) {#1};}
\newcommand{\lLL}[1]{\node[font=\scriptsize] at ($(L)!0.5!(BR)+(-0.27,-0.17)$) {#1};}
\newcommand{\triMbb}[2]{\begin{feynman}\diagram*{(L)--[#1](TR),(TR)--[#2](BR),(L)--[fermion](BR),};\end{feynman}}
\newcommand{\triMbB}[2]{\begin{feynman}\diagram*{(L)--[#1](TR),(TR)--[#2](BR),};\end{feynman}\draw[dblplain](L)--(BR);}
\newcommand{\triBbb}[1]{\begin{feynman}\diagram*{(L)--[fermion](TR),(TR)--[fermion](BR),(L)--[#1](BR),};\end{feynman}}
\newcommand{\triBBb}[1]{\begin{feynman}\diagram*{(TR)--[fermion](BR),(L)--[#1](BR),};\end{feynman}\draw[dblplain](L)--(TR);}
\newcommand{\triBbB}[1]{\begin{feynman}\diagram*{(L)--[fermion](TR),(L)--[#1](BR),};\end{feynman}\draw[dblplain](TR)--(BR);}
\newcommand{\triBBB}[1]{\begin{feynman}\diagram*{(L)--[#1](BR),};\end{feynman}\draw[dblplain](L)--(TR);\draw[dblplain](TR)--(BR);}

\begin{document}

\title{Radiative decays of dynamically generated pentaquarks in the chiral
unitary approach: the $P_c(4457)\to P_c(4312)\,\gamma$ transition}

\author{Ratirat Suntharawirat}
\email{ratirat_su@kkumail.com}
\author{Nongnaphat Ponkhuha}
\email{Nongnapat.po@kkumail.com}
\author{Daris Samart}
\thanks{Corresponding author.}
\email{darisa@kku.ac.th}
\affiliation{Khon Kaen Particle Physics and Cosmology Theory Group (KKPaCT),
Department of Physics, Faculty of Science, Khon Kaen University,
123 Mitraphap Rd., Khon Kaen 40002, Thailand}

\date{\today}

\begin{abstract}
{\textcolor{black}{We study the radiative decay of dynamically generated pentaquarks and apply the formalism to the transition
$P_c(4457)(3/2^-)\to P_c(4312)(1/2^-)\gamma$. Both states are treated as $S$-wave hadronic molecules generated in the chiral unitary approach with heavy-quark spin symmetry and the local hidden gauge interaction. The photon therefore couples to the meson-baryon components of the two poles. The calculation combines the strong coupling residues of the coupled-channel solution, heavy-quark spin symmetry for the electromagnetic vertices, and a transverse assembly of the $M1$ triangle loops. A complete calculation gives nineteen triangle loops. We reduce each loop to a single numerical quadrature and give the closed analytic form. The electromagnetic vertices that are not fixed by data are estimated with the naive-quark model and heavy-quark spin symmetry. We normalize the main $D^*D\gamma$ coupling to $\bar D^{*0}\to\bar D^{0}\gamma$ and test the same convention with $J/\psi\to\eta_c\gamma$. The central width is $6.7\keV$, with a conservative range of about $2$ to $9\keV$. This radiative decay process is a pure $M1$ transition with photon energy $143\MeV$. The $\bar D^{*0}\to\bar D^{0}\gamma$ loop gives the leading contribution. The near-threshold $\bar D^{*}\Lambda_c$ loop gives the main correction. \textcolor{black}{A soft Gaussian form-factor on the leading diagram reduces the width to about $2\keV$, compatible with earlier molecular results, and decreases the full width to about $4\keV$.} The coherent result is sensitive to the relative residue phases in the coupled-channel convention. We also estimate the cascade rate for $\Lambda_b^{0}\to J/\psi\,p\,K^{-}\gamma$ and discuss how the line can be searched for. \textcolor{black}{The pure $M1$ content, the ratio to the $P_c(4440)$ radiative decay, and the binding-energy dependence of the width are proposed as tests of the molecular nature.}}}
\end{abstract}


\maketitle

%=====================================================================
\section{Introduction}
\label{sec:intro}

The LHCb collaboration observed three narrow pentaquark states in the $J/\psi\,p$ spectrum of $\Lambda_b^{0}\to J/\psi\,p\,K^{-}$ decays \cite{LHCb2019}. These are the $P_c(4312)$, the $P_c(4440)$ and the $P_c(4457)$. An earlier analysis had reported pentaquark candidates in the same final state \cite{LHCb2015}. The three peaks sit a few MeV below the $\Db\Sc$ and $\Dbs\Sc$ thresholds. {\textcolor{black}{This threshold pattern indicates a molecular interpretation of the states.}}

In the chiral unitary and local hidden gauge coupled-channel approach the three states come out as $S$-wave quasi-bound states of a charmed meson and a charmed baryon \cite{XNO2013,XNO2019}. The chiral unitary approach itself, which combines chiral perturbation theory with $N/D$ or Bethe-Salpeter
re-summation in coupled channels, was developed for the light meson and meson-baryon sectors in
Refs.~\cite{KaiserSiegelWeise1995,OllerOset1997,OsetRamos1998,OllerOsetPelaez1999}. The same idea had predicted hidden-charm baryons above $4\GeV$ before the data \cite{WuMolinaOsetZou2010,WuMolinaOsetZou2011,HofmannLutz2005}. Other groups reached the same molecular conclusion using one-boson-exchange and effective field theory frameworks
\cite{KarlinerRosner2015,ChenLiuLiZhu2015,He2016,LiuPanPengEtAl2019,ChenSunLiuZhu2019,HeQuasipotential2019,DuBaruGuoHanhartEtAl2020,GuoOller2019,RocaNievesOset2015},
and the molecular picture is reviewed in
Refs.~\cite{ChenReview2016,GuoHadronicMolecules2018,LiuChenChenLiuZhu2019,BrambillaEtAl2020,ChenLiuLiuLiuZhu2023}.
In the solution of Refs.~\cite{XNO2013,XNO2019}, the $P_c(4312)$ is mostly $\Db\Sc$ with $J^{P}=\half^{-}$ while the $P_c(4440)$ and $P_c(4457)$ are mostly $\Dbs\Sc$ with $J^{P}=\tfrac12^{-}$ and $\tfrac32^{-}$, respectively.

The two states share the same $\Sc$ core. They differ mainly in the spin of the charmed meson. An $M1$ photon transition between them is a direct test of their inner structure. A molecular state radiates through its meson and baryon parts. A compact state, in the diquark-diquark-antiquark or hadrocharmonium picture
\cite{MaianiPolosaRiquer2015,Lebed2015,Wang2016SumRules}, radiates through its quark core. Cusp and triangle-singularity readings of the LHCb peaks have also been put forward \cite{GuoMeissnerWangYang2015,BayarAcetiGuoOset2016}. The three pictures give different rates. The photon here is soft, with an energy near $143\MeV$, yielding a signal well-separated from background noise.

Electromagnetic decays of the $P_c$ states have been studied in several frameworks. In the quark model, the photon couples to the constituent quarks. In this picture the magnetic moments and the photocouplings of the $P_c$ states were obtained \cite{WangChenMaLiuZhu2016,OrtizBijkerFernandez2019}. In the
phenomenological Lagrangian and molecular picture the decay properties of the $P_c$ states were worked out, including the magnetic moments and the transition magnetic moments between them
\cite{WangXiaoChen2020,LiLiuSunChen2021}. The $M1$ radiative widths and the magnetic moments of related molecular pentaquarks were also computed in the same molecular framework \cite{LaiWangLiu2024}. {\textcolor{black}{The same process
$P_c(4457)\to P_c(4312)\gamma$ was studied in two earlier works. Ref.~\cite{LingLuLiuGeng2021} used an effective Lagrangian and a single $\bar D^{*}\to\bar D\gamma$ triangle and found a width \textcolor{black}{of about $1.4$ to $2.2\keV$}. Ref.~\cite{LiLiuSunChen2021} studied transition magnetic moments and found $1.1$ to $1.7\keV$. Both keep the leading meson radiation only. We compare with these two results in Sec.~\ref{sec:width}. \textcolor{black}{We also compare the leading diagram under the same form-factor in Sec.~\ref{sec:formfactor}.}}} In QCD sum rules the
electromagnetic multipole moments were extracted, both for the $P_c(4380)$ in
light-cone QCD \cite{OzdemAzizi2018} and for the $P_c(4440)$, the $P_c(4457)$ and the $P_{cs}(4459)$ \cite{Ozdem2021}, and the magnetic moment of the $P_c(4312)$ was treated as a $\Db\Sc$ molecular state \cite{XuLiuHuang2021}. The coupled-channel route, in which the photon couples to the full meson-baryon content of the state, is the one we follow here.

Most of these works give a static magnetic moment or a single transition magnetic moment. A systematic and gauge-invariant treatment of the radiative decays of the pentaquarks as dynamically generated states is still missing. We provide it here, and we apply it to the inter-state transition $P_c(4457)\to P_c(4312)\gamma$. We compute the width in the coupled-channel approach with the complete set of Ref.~\cite{XNO2019} residues. We keep every meson-baryon pair of the two poles. Gauge invariance then asks for nineteen $M1$ triangle loops.

The construction brings together three ingredients that have not been combined before for the pentaquark radiative sector. First, the two states are dynamically generated in the chiral unitary approach, so the photon couples to the full coupled-channel content of each pole through its residues. Second, heavy-quark spin symmetry fixes the strong molecular vertices and the
heavy-meson radiative vertices on the same footing, and the naive-quark model fixes the light-quark moments. Third, the local hidden gauge interaction and a gauge invariant assembly of the triangle loops give a transverse $M1$ amplitude that one can add loop by loop. The outcome is a complete and gauge invariant sum
over the nineteen meson-baryon triangles of the two poles, with the complex residue phases kept throughout. To our knowledge this is the first such calculation for the pentaquark radiative decays. It also serves as a template. The same scheme carries over to the partner transition $P_c(4440)\to P_c(4312)\gamma$ and to the strange sector without introducing new  parameter.

The chiral unitary scheme has a clear advantage for this problem. A dynamically generated resonance is a meson-baryon composite. The photon couples to each meson-baryon component in turn, through long-distance light-quark moments. The rate reads the molecular wave function directly. This differs from the quark model, where in the heavy-quark limit
\cite{IsgurWise1989,IsgurWise1990,Neubert1994,ManoharWise2000} the photon couples only to the light diquark of a compact core \cite{GamermannBaryon2011}. The chiral unitary approach has been successfully applied to other exotic hadrons. The Valencia group computed the radiative decay of the $\Lambda(1520)$ as a dynamically generated $\tfrac32^{-}$ state \cite{DoringOsetSarkar2006}, the radiative decays of resonances generated from the vector-baryon interaction
\cite{SunGarzonOset2010}, the radiative decay of the open and hidden charm scalar mesons $D_{s0}^{*}(2317)$ and $X(3700)$ \cite{GamermannMeson2007}, and the radiative decays of dynamically generated charmed baryons \cite{GamermannBaryon2011}. We follow this template. We construct gauge invariant triangle loops from the same propagators and the same coupled channels that generate the two poles.

The paper is organized as follows. Sec.~\ref{sec:formalism} gives the formalism. Sec.~\ref{sec:cua} recalls the chiral unitary approach and the two poles. Sec.~\ref{sec:rules} lists the propagators and the non-relativistic hadronic vertices, the electromagnetic vertices, and the structure of the radiative amplitude. Sec.~\ref{sec:amps} sets up the nineteen loop amplitudes, first as loop integrals and then in their final one-parameter form. Sec.~\ref{sec:results} gives the numerical width and its physical implications, and then
the rate of the $\Lambda_b^{0}\to J/\psi\,p\,K^{-}\gamma$ cascade.
Sec.~\ref{sec:conclusions} is the discussion and the conclusions. %The appendices collect the spin algebra, the loop master formulas, the width formula, and the closed analytic form of every amplitude.

%=====================================================================
\section{Formalism}
\label{sec:formalism}

%---------------------------------------------------------------------
\subsection{Chiral unitary approach}
\label{sec:cua}
%%%-----------------------------------------------------------------------------------------------------------------------------------
\begin{table}[b]
\centering
\renewcommand{\arraystretch}{1.25}
\begin{tabular}{lcc}
\hline\hline
 & $P_c(4312)$ & $P_c(4457)$\\
channel & $J^{P}=\half^{-}$ & $J^{P}=\tfrac32^{-}$\\
\hline
$\eta_c N$ & $0.67{+}0.01i$ & \\
$J/\psi N$ & $0.46{-}0.03i$ & $0.30{-}0.01i$\\
$\Db\Lc$   & $0.01{-}0.01i$ & \\
$\Db\Sc$   & $2.07{-}0.28i$ & \\
$\Db\Scs$  &                & $0.08{-}0.02i$\\
$\Dbs\Lc$  & $0.03{+}0.25i$ & $0.05{-}0.04i$\\
$\Dbs\Sc$  & $0.06{-}0.31i$ & $1.82{-}0.08i$\\
$\Dbs\Scs$ & $0.04{-}0.15i$ & $0.01{-}0.19i$\\
\hline\hline
\end{tabular}
\caption{Dimensionless residues of the two poles in the chiral unitary
solution of Ref.~\cite{XNO2019}, in the isospin basis. We use the shorthands
$g_1\equiv g_{\Pcp\Dbs\Sc}=1.82-0.08i$, $g_2\equiv g_{\Pc\Db\Sc}=2.07-0.28i$
and $g_3\equiv g_{\Pc\Dbs\Sc}=0.06-0.31i$.}
\label{tab:residues}
\end{table}
%%%-----------------------------------------------------------------------------------------------------------------------------------
The coupled-channel amplitude is $T=[1-VG]^{-1}V$. The kernel $V$ is the local hidden gauge interaction \cite{HiddenGauge1988} extended to four flavors with heavy-quark spin symmetry
\cite{IsgurWise1989,IsgurWise1990,Neubert1994,ManoharWise2000,XNO2013}. The function $G$ is the diagonal meson-baryon two-point loop function. For a channel with baryon mass $M_l$ and meson mass $m_l$ it reads
\begin{equation}
G_l(s)=i\!\int\!\frac{d^4q}{(2\pi)^4}\;
\frac{2M_l}{(P-q)^{2}-M_l^{2}+i\eps}\;\frac{1}{q^{2}-m_l^{2}+i\eps}.
\label{eq:Gdef}
\end{equation}
We regularize $G_l$ in dimensional regularization. The subtraction constant is $a_l(\mu)=-2.09$ at $\mu=1\GeV$, as in the model of Ref.~\cite{XNO2019}. A pole $\sqrt{s_0}$ in the second sheet defines a state. Near the pole the amplitude factorizes,
\begin{equation}
T_{ij}\simeq \frac{g_i\,g_j}{\sqrt{s}-\sqrt{s_0}}.
\label{eq:residue}
\end{equation}
The residues $g_i$ are the couplings of the state to its channels. They do not carry dimension in this normalization. The pole positions are $\sqrt{s_0}=4306.4+i\,7.6\MeV$ for $P_c(4312)$ and $\sqrt{s_0}=4452.5+i\,1.5\MeV$ for $P_c(4457)$.

Tab.~\ref{tab:residues} lists the residues of the two poles \cite{XNO2019}. The $P_c(4312)$ has seven channels. The $P_c(4457)$ has five. A pseudoscalar meson with a $\half^{+}$ baryon cannot form $\tfrac32^{-}$ in the $S$-wave, so the
$\tfrac32^{-}$ pole has fewer channels. We write $g_1$ for the residue $\Dbs\Sc$ of $P_c(4457)$, $g_2$ for the residue $\Db\Sc$ of $P_c(4312)$, and $g_3$ for its residue $\Dbs\Sc$. The residue $g_2$ is close to real. {\textcolor{black}{The residues $g_3$ and the $\Dbs\Lc$ residue are mostly imaginary in this convention. Their relative phases, once combined with the electromagnetic matrix elements in the same convention, control the interference pattern of the radiative amplitude. Only such convention-consistent products are observable.}} %\textcolor{black}{The spin assignments $J^{P}(P_c(4457))=\tfrac32^{-}$ and $J^{P}(P_c(4312))=\tfrac12^{-}$ are taken from the coupled-channel solution. The LHCb data do not fix them. The rate and the pure $M1$ structure depend on this choice.}
More importantly, the assignments \(J^{P}=\tfrac32^{-}\) for the \(P_c(4457)\) and \(J^{P}=\tfrac12^{-}\) for the \(P_c(4312)\) are adopted from the coupled-channel results, as they are not determined by the LHCb data. Consequently, both the decay rate and the pure \(M1\) structure depend on this choice.

The radiative transition replaces one propagator of $G_l$ with two propagators joined by a photon vertex. {\textcolor{black}{This gives a triangle diagram.}} The triangle uses the same
propagators and the same $2M_B$ baryon normalization as $G_l$, so the residues enter without additional factor. We write $M_1$ for the mass of the line that enters the photon vertex, $M_2$ for the line after it, and $M_s$ for the spectator.
The scalar triangle is
\begin{widetext}
\begin{equation}
\mathcal I=i\!\int\!\frac{d^4q}{(2\pi)^4}\;
\frac{(2M_B)^{n_B}\,[\text{numerator}]}
{[q^{2}-M_1^{2}+i\eps]\,[(q-K')^{2}-M_2^{2}+i\eps]\,[(P-q)^{2}-M_s^{2}+i\eps]},
\label{eq:tridef}
\end{equation}
\end{widetext}
with $n_B$ the number of baryon lines and $K'$ the photon momentum routed onto
the radiating line. Sec.~\ref{sec:amps} reduces Eq.~\eqref{eq:tridef} to a two-parameter Feynman integral over $x$ and $y$. The closed form of the inner integral is given in \ref{app:yint}.

%---------------------------------------------------------------------
\subsection{Feynman rules and the radiative amplitude}
\label{sec:rules}

We work in the rest frame of the initial state,
\begin{equation}
\begin{gathered}
\Pcp(P)\to\Pc(Q)+\gamma(K),\\
P=(M_i,\bm 0),\quad K=(E_\gamma,\bm K),\quad Q=P-K,
\end{gathered}
\end{equation}
with $K^2=0$ and $E_\gamma=(M_i^{2}-M_f^{2})/2M_i$. The masses
$M_i=4457.3\MeV$ and $M_f=4311.9\MeV$ give $E_\gamma=143.0\MeV$. We use Coulomb gauge, $\eps^{0}=0$ and $\bm K\cdot\beps=0$, with
$\sum_\lambda\eps_i\eps_j^{*}=\delta_{ij}-\hat K_i\hat K_j$. Every photon vertex we keep is magnetic, so the amplitudes are transverse and this gauge is a convenience. %The photon is external, so its propagator never appears.

\paragraph{Fields and spin operators.}
All hadrons in the loop are heavy and near their mass shells, so we use non-relativistic normalizations. Spin-$\half$ baryons are two-component Pauli spinors $\chi$ with $\chi^\dagger\chi=1$. The spin-$\tfrac32$ $\Pcp$ and $\Scs$ are described by $\half\to\tfrac32$ transition-spin operators $\bm S$ (and
$\bm S_B$ for $\Scs$), with
\begin{equation}
S_i S_j^\dagger=\tfrac23\,\delta_{ij}-\tfrac{i}{3}\,\varepsilon_{ijk}\sigma_k,
\qquad \bm S\cdot\bm S^\dagger=2,
\label{eq:SSdag}
\end{equation}
and the identities, proved in \ref{app:spin},
\begin{equation}
\varepsilon_{jkm}\,\sigma_m S_j=-\,i\,S_k,
\qquad
\sum_k S_k\,(\bm S^\dagger\!\cdot\bm A)\,S_k=\tfrac13\,\bm A\cdot\bm S.
\label{eq:keyid}
\end{equation}

\paragraph{Propagators.}
The mesons keep relativistic propagators and the baryon lines carry the
normalization in which the coupled-channel couplings are dimensionless. For a
pseudoscalar meson ($\Db,\eta_c$),
\begin{equation}
i\Delta_P(q)=\frac{i}{q^{2}-m_P^{2}+i\eps}.
\label{eq:propP}
\end{equation}
For a vector meson ($\Dbs,J/\psi$),
\begin{equation}
\begin{aligned}
i\Delta_V^{\mu\nu}(q)=\frac{i\big(-g^{\mu\nu}+\frac{q^\mu q^\nu}{m_V^{2}}\big)}
{q^{2}-m_V^{2}+i\eps}\xrightarrow[\text{NR}]{}\;
i\Delta_V^{ij}(q)=\frac{i\,\delta^{ij}}{q^{2}-m_V^{2}+i\eps},
\end{aligned}
\label{eq:propV}
\end{equation}
the $q^iq^j/m_V^{2}$ part being dropped as $\mathcal O(\bm q^{2}/m_V^{2})$ with
the time components. \textcolor{black}{The loop momentum is set by the regulator and the binding of the molecular states, of order a few hundred $\MeV$, so $\bm q^{2}/m_V^{2}$ is at the percent level and this part is a small correction.} For a spin-$\half$ baryon ($N,\Lc,\Sc$) the relativistic propagator reduces the positive-energy projection to
\begin{equation}
iS_B(q)=\frac{i\,(2M_B)}{q^{2}-M_B^{2}+i\eps}\,\mathbf 1_{2\times2},
\label{eq:propB}
\end{equation}
the factor $2M_B=\bar u u$ being the normalization that converts the relativistic line to the loop function $G_l$. For a spin-$\tfrac32$ baryon ($\Scs$) the Rarita-Schwinger propagator reduces to
\begin{equation}
\begin{gathered}
iS^{ij}(q)=\frac{i\,(2M)}{q^{2}-M^{2}+i\eps}\,P^{ij}_{3/2},\\
P^{ij}_{3/2}=\delta^{ij}-\tfrac13\sigma^i\sigma^j
=\sum_{m}S_B^{i}\,|m\rangle\langle m|\,S_B^{j\dagger},
\end{gathered}
\label{eq:propBs}
\end{equation}
then the spin-$\tfrac32$ projector is realized by the transition operators of Eq.~\eqref{eq:SSdag}.

\paragraph{Isospin factor.}
With $\Db=(\Db^{0},D^{-})$ and $\Sc=(\Sc^{++},\Sc^{+},\Sc^{0})$, the
$I=\half$, $I_3=+\half$ combination is
\begin{equation}
\big|\Db^{(*)}\Sc;\tfrac12,+\tfrac12\big\rangle
=\sqrt{\tfrac13}\,\big|\Db^{(*)0}\Sc^{+}\big\rangle
-\sqrt{\tfrac23}\,\big|D^{(*)-}\Sc^{++}\big\rangle.
\label{eq:isospin}
\end{equation}
When the photon does not change the meson-baryon charge the same charge channel
appears at both molecular vertices, giving the weights
$C_{\rm n}^{2}=\tfrac13$ (neutral) and $C_{\rm c}^{2}=\tfrac23$ (charged). The
$\Db^{(*)}\Lc$ channels have an $I=0$ baryon. A $\Sc\to\Lc$ transition vertex
then projects $I=1$ onto $I=0$ and brings a factor $\sqrt{1/3}$. For $J/\psi N$
and $\eta_c N$ there is one charge state and the weight is unity.

\paragraph{Hadronic vertices.}
With the residues of Tab.~\ref{tab:residues}, the non-relativistic vertex
functions, per charge channel and before the isospin factors, are given by
\begin{widetext}
\begin{align}
\Pcp(\tfrac32^-)\to V(\eps_V)\,B(\tfrac12) &\;\Rightarrow\;
-i\,g\;\chi_{B}^\dagger\,(\bm S\cdot\beps_V^{\,*})\,\chi_{\Pcp}
&& [\Dbs\Sc,\,\Dbs\Lc,\,J/\psi N],
\label{eq:v32V}\\
\Pcp(\tfrac32^-)\to P\,B^{*}(\tfrac32) &\;\Rightarrow\;
-i\,g\;\chi_{B^{*}}^{\dagger}\,\chi_{\Pcp}
&& [\Db\Scs],
\label{eq:v32P}\\
\Pcp(\tfrac32^-)\to V(\eps_V)\,B^{*}(\tfrac32) &\;\Rightarrow\;
-i\,g\;\chi_{B^{*}}^{\dagger}\,\bm S^{\dagger}\cdot\,(\bm\sigma\cdot\beps_V^{\,*})\,\bm S\,\chi_{\Pcp}
&& [\Dbs\Scs],
\label{eq:v32Vs}\\
\Pc(\tfrac12^-)\to P\,B(\tfrac12) &\;\Rightarrow\;
-i\,g\;\chi_{B}^\dagger\,\chi_{\Pc}
&& [\Db\Sc,\,\Db\Lc,\,\eta_c N],
\label{eq:v12P}\\
\Pc(\tfrac12^-)\to V(\eps_V)\,B(\tfrac12) &\;\Rightarrow\;
-i\,\frac{g}{\sqrt3}\;\chi_{B}^\dagger\,(\bm\sigma\cdot\beps_V^{\,*})\,\chi_{\Pc}
&& [\Dbs\Sc,\,\Dbs\Lc,\,J/\psi N],
\label{eq:v12V}\\
\Pc(\tfrac12^-)\to V(\eps_V)\,B^{*}(\tfrac32) &\;\Rightarrow\;
-i\,\frac{g}{\sqrt2}\;\chi_{B^{*}}^{\dagger}\,(\bm S_B^\dagger\!\cdot\beps_V^{\,*})\,\chi_{\Pc}
&& [\Dbs\Scs].
\label{eq:v12Vs}
\end{align}
\end{widetext}
The $1/\sqrt3$ in Eq.~\eqref{eq:v12V} restores the $J=\half$ normalization of
the Cartesian vertex, and the $1/\sqrt2$ in Eq.~\eqref{eq:v12Vs} is the analogue
for a $\half\to VB^{*}(\tfrac32)$ vertex. The scalar and $\bm S\cdot\beps$
vertices need no such factor. In Eq.~\eqref{eq:v32Vs} both the $\Pcp$ and the
$\Scs$ carry spin $\tfrac32$, so the vector-meson spin $\bm\sigma$ is recoupled
between the two transition operators. This is the structure we mark as a spin-2
recoupling below.

Every vertex in Eqs.~\eqref{eq:v32V} to \eqref{eq:v12Vs} is an $S$-wave
coupling. It is allowed because in each channel
$\eta_M\eta_B=-1=\eta_{P_c^{(\prime)}}$, so no derivative coupling is needed.
The parity of each operator and its fields is checked in
Tab.~\ref{tab:parity}. Every line gives $+1$, so all six vertices are parity
conserving as written.

\begin{table*}[tb]
\centering
\renewcommand{\arraystretch}{1.3}
\setlength{\tabcolsep}{5pt}
\begin{tabular}{clccccc}
\toprule
Eq. & vertex $P_c^{(\prime)}\!\to\! M\,B$ & $J^P_M$ & $J^P_B$ &
$\eta_M\eta_B$ & operator $\mathcal O$ & $\mathsf P[\,\mathcal O\,\&\,\text{fields}\,]$\\
\midrule
\eqref{eq:v32V}  & $\Pcp\!\to\! V\,B(\tfrac12)$       & $1^-$ & $\tfrac12^+$ & $-$ & $\bm S\!\cdot\!\beps_V^{*}$ & $+$\\
\eqref{eq:v32P}  & $\Pcp\!\to\! P\,B^{*}(\tfrac32)$   & $0^-$ & $\tfrac32^+$ & $-$ & $\mathbf 1$  & $+$\\
\eqref{eq:v32Vs} & $\Pcp\!\to\! V\,B^{*}(\tfrac32)$   & $1^-$ & $\tfrac32^+$ & $-$ & $\bm S^{\dagger}\cdot(\bm\sigma\cdot \bm\eps_V^{*})\, \bm S$ & $+$\\
\eqref{eq:v12P}  & $\Pc\!\to\! P\,B(\tfrac12)$        & $0^-$ & $\tfrac12^+$ & $-$ & $\mathbf 1$ & $+$\\
\eqref{eq:v12V}  & $\Pc\!\to\! V\,B(\tfrac12)$        & $1^-$ & $\tfrac12^+$ & $-$ & $\bm\sigma\!\cdot\!\beps_V^{*}$ & $+$\\
\eqref{eq:v12Vs} & $\Pc\!\to\! V\,B^{*}(\tfrac32)$    & $1^-$ & $\tfrac32^+$ & $-$ & $\bm S_B^{\dagger}\!\cdot\!\beps_V^{*}$ & $+$\\
\bottomrule
\end{tabular}
\caption{Parity audit of the six molecular vertices. The last column is the
product of the parity eigenvalues of the spin operator and of all participating
fields, including the $1^-$ polarization flip $\beps_V\!\to\!-\beps_V$ and the
$-1$ of the negative-parity $\chi_{P_c^{(\prime)}}$. It equals $+1$ for every
vertex.}
\label{tab:parity}
\end{table*}

\paragraph{Electromagnetic vertices.}
%Data fix the $D^{*}\to D\gamma$ transitions. 
We define the coupling $\lambda$ by the standard $V P\gamma$ vertex of the heavy-hadron chiral Lagrangian
\cite{Wise1992,YanChengChiangChiangChiangYu1992,AmundsonBoydJenkinsLukeManoharPolitzerWise1992,ChoWise1994,CasalbuoniDeandreaDiBartolomeoGattoFeruglioNardulli1997},
\begin{equation}
\begin{gathered}
-i\,t_{V P\gamma}= -i\,e\,\lambda\,(\bK\times\beps_\gamma^{\,*})\cdot\beps_V,\\
\Gamma(D^{*}\to D\gamma)=\frac{\alpha}{3}\,\lambda^{2}\,E_\gamma^{\prime\,3}.
\end{gathered}
\label{eq:lamdef}
\end{equation}
The measured widths $\Gamma(D^{*0}\to D^{0}\gamma)=19.5\keV$ and
$\Gamma(D^{*+}\to D^{+}\gamma)=1.33\keV$ \cite{PDG2024} give
\begin{equation}
\begin{gathered}
\lambda_0=\lambda(\bar D^{*0}\Db^{0}\gamma)=+1.766\GeV^{-1},\\
\lambda_-=\lambda(D^{*-}D^{-}\gamma)=-0.469\GeV^{-1}.
\end{gathered}
\label{eq:lambdas}
\end{equation}
The other vertices follow in the naive-quark model with heavy-quark spin symmetry \cite{IsgurWise1989,Neubert1994,ManoharWise2000}. An $M1$ photon flips one constituent spin with a strength set by its magnetic moment. We use
$\mu_u=1.852$, $\mu_d=-0.972$ and $\mu_c=0.404$ in nuclear magnetons, the constituent values obtained from baryon magnetic moments and quoted in Ref.~\cite{PDG2024}. In a $q\bar c$ meson the $V\to P$ transition moment is the difference of the two constituent moments, $\mu_q-\mu_{\bar c}=\mu_q+\mu_c$, since the antiquark moment enters with the opposite sign. This sets one
constant,
\begin{equation}
\begin{gathered}
\lambda_0=N(\mu_u+\mu_c),\qquad
\lambda_-=N(\mu_d+\mu_c),\\
N=\frac{\lambda_0}{\mu_u+\mu_c}=0.783\GeV^{-1}/\mu_N.
\end{gathered}
\label{eq:Ncalib}
\end{equation}
As a check, $N(\mu_d+\mu_c)=-0.445\GeV^{-1}$ agrees with the data value $\lambda_-=-0.469\GeV^{-1}$ at the five percent level. The elastic $\Dbs$ moment is the sum $\mu_q-\mu_c$ instead, because no spin flips there. The same flip rule gives $\lambda(\psi\eta_c\gamma)=2N\mu_c=0.633\GeV^{-1}$. This yields
$\Gamma(J/\psi\to\eta_c\gamma)=1.33\keV$. \textcolor{black}{The measured decay width is not fixed. The PDG 2024 average branching fraction gives about $1.3\keV$. The PDG fit gives about $1.7\keV$ \cite{PDG2024}. The direct measurements run from about $1.2\keV$ to about $2\keV$. Our value sits at the low end of this range. %We take this spread as the accuracy of the naive-quark model $M1$ rule in the heavy quark sector. We fold it into the systematic uncertainty budget.
This spread is taken as a measure of the accuracy of the naive-quark model \(M1\) rule in the heavy-quark sector and is incorporated directly into the systematic uncertainty budget.} %\textcolor{black}{The prediction is lower about twenty percent. This fixes the accuracy of the naive-quark-model $M1$ rule in the heavy sector and is folded into the uncertainty budget.}

The baryon vertices follow the same manner. The $\Sc^{+}\to\Lc^{+}\gamma$ transition plays the major role. The light diquark changes spin from one to zero, like $\Sigma^{0}\to\Lambda\gamma$. We find
$\mu(\Sc^{+}\to\Lc^{+})=-(\mu_u-\mu_d)/\sqrt3=-1.63\,\mu_N$, equal in size to the measured $\mu(\Sigma^{0}\Lambda)$. The spin-$\tfrac32$ partner is a factor $\sqrt2$ larger. The sign of these baryon transition moments relative to the
residue basis is not fixed by the quark model. This sign is the main source of
uncertainty in the width. The transition vertices read
$-i\,e\,\tilde\mu\,\chi^\dagger\,\bm\sigma\cdot(\bK\times\beps_\gamma^{\,*})\,\chi$
for spin-$\half$ baryons and
$-i\,e\,\tilde\mu_3\,\chi^\dagger\,\bm S_B^\dagger\cdot(\bK\times\beps_\gamma^{\,*})\,\chi$
for the $\Scs\to B$ transitions, with $\tilde\mu=\mu/2m_N$.

\textcolor{black}{Three further vertex families enter the loops. The elastic vector vertex $V\to V\gamma$ sits on the $\Dbs$ and the $J/\psi$ lines. It carries the magnetic coupling $\lambda_g$ and flips no spin, so its moment is the sum \textcolor{black}{$\mu_q+\mu_{\bar c}$, equal to $\mu_q-\mu_c$ with the antiquark moment $\mu_{\bar c}=-\mu_c$}. By $C$ parity the $J/\psi$ has no such moment, so its elastic
vertex is absent. The baryon transition vertex $B^{*}\to B\gamma$ sits on the $\Scs\to\Sc$ and $\Scs\to\Lc$ lines with the magnetic moment $\tilde\mu_3$. The elastic vertex $B^{*}\to B^{*}\gamma$ sits on the $\Scs$ line with the moment $\tilde\mu_{B^{*}}$. The three forms are the baryon mirror of $V\to P\gamma$ and
$V\to V\gamma$. Each one is a pure $M1$ structure proportional to
$\bK\times\beps_\gamma^{\,*}$, so every triangle reduces to the universal structure of Eq.~\eqref{eq:Mstructure}. The explicit vertices, the quark-model value of every magnetic moment, and the proof that the electric part of each elastic vertex drops out of the loop are given in App.~\ref{app:em}.}
%%%%%%%%%%%%%%%%%%%%%%%%%%%%%%%%%%%%%%%%%%%%%%%%%%%%%%%%%%%%%%%%%%%%%%%%%%%%%%%%%%%%%%%%%%%%%%%%%%%%%%%%%%%%%%%%%%%%%%%%%%%%%%%%%%%%%%%%%%%%%%%%
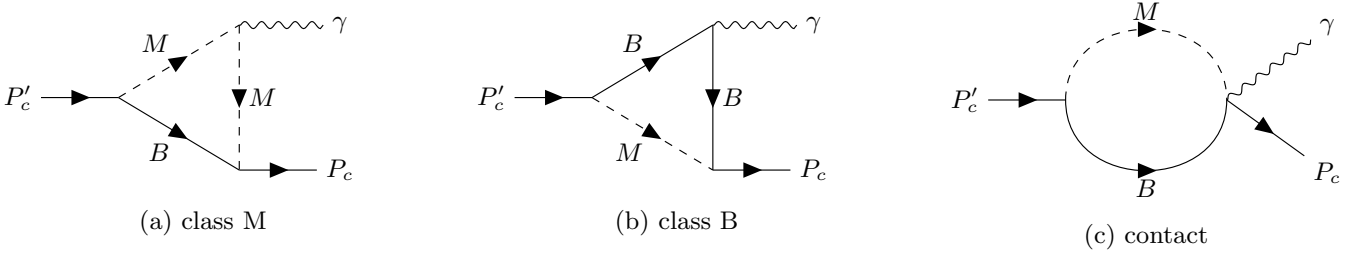
\begin{figure*}[t]
\centering
\resizebox{\textwidth}{!}{%
\begin{tikzpicture}[baseline=(current bounding box.center)]
\begin{feynman}
\vertex (in) at (-1.25,0) {\(\Pcp\)};
\vertex (L)  at (0,0);
\vertex (tr) at (1.5,0.9);
\vertex (br) at (1.5,-0.9);
\vertex (g)  at (2.75,0.9) {\(\gamma\)};
\vertex (f)  at (2.75,-0.9) {\(\Pc\)};
\diagram*{
 (in) -- [fermion] (L),
 (L)  -- [charged scalar, edge label=\(M\)] (tr),
 (tr) -- [charged scalar, edge label=\(M\)] (br),
 (L)  -- [fermion, edge label'=\(B\)] (br),
 (tr) -- [photon] (g),
 (br) -- [fermion] (f),
};
\end{feynman}
\node at (1.05,-1.55) {(a) class M};
\end{tikzpicture}
\hspace{1.1cm}
\begin{tikzpicture}[baseline=(current bounding box.center)]
\begin{feynman}
\vertex (in) at (-1.25,0) {\(\Pcp\)};
\vertex (L)  at (0,0);
\vertex (tr) at (1.5,0.9);
\vertex (br) at (1.5,-0.9);
\vertex (g)  at (2.75,0.9) {\(\gamma\)};
\vertex (f)  at (2.75,-0.9) {\(\Pc\)};
\diagram*{
 (in) -- [fermion] (L),
 (L)  -- [fermion, edge label=\(B\)] (tr),
 (tr) -- [fermion, edge label=\(B\)] (br),
 (L)  -- [charged scalar, edge label'=\(M\)] (br),
 (tr) -- [photon] (g),
 (br) -- [fermion] (f),
};
\end{feynman}
\node at (1.05,-1.55) {(b) class B};
\end{tikzpicture}
\hspace{1.1cm}
\begin{tikzpicture}[baseline=(current bounding box.center)]
\begin{feynman}
\vertex (in) at (-1.25,0) {\(\Pcp\)};
\vertex (L)  at (0,0);
\vertex (R)  at (2.0,0);
\vertex (g)  at (3.25,0.9) {\(\gamma\)};
\vertex (f)  at (3.25,-0.9) {\(\Pc\)};
\diagram*{
 (in) -- [fermion] (L),
 (L)  -- [charged scalar, half left, edge label=\(M\)] (R),
 (L)  -- [fermion, half right, edge label'=\(B\)] (R),
 (R)  -- [photon] (g),
 (R)  -- [fermion] (f),
};
\end{feynman}
\node at (1.0,-1.7) {(c) contact};
\end{tikzpicture}}
\caption{The two triangle topologies and the contact topology. In class M~(a) the photon comes off a meson line $M$ and a baryon $B$ is the spectator. In class B~(b) the photon comes off a baryon line $B$ and a meson $M$ is the spectator. \textcolor{black}{Diagram~(c) is the contact, or Kroll-Ruderman term, where the photon is emitted at the final $P_cMB$ vertex. It vanishes
for the $S$-wave molecular vertices used here, as shown in
Sec.~\ref{sec:rules} and App.~\ref{app:gauge-conv}, so it is  not included to loop amplitudes.} The $P_c(4457)$ enters on the left and the $P_c(4312)$ leaves on the right.}
\label{fig:topology}
\end{figure*}
%%%%%%%%%%%%%%%%%%%%%%%%%%%%%%%%%%%%%%%%%%%%%%%%%%%%%%%%%%%%%%%%%%%%%%%%%%%%%%%%%%%%%%%%%%%%%%%%%%%%%%%%%%%%%%%%%%%%%%%%%%%%%%%%%%%%%%%%%%%%%%%%
\paragraph{Structure of the amplitude.}
For $\tfrac32^{-}\to\half^{-}\gamma$ only $M1$ and $E2$ are allowed. All hadronic vertices are $S$-wave and momentum independent. The magnetic photon vertex gives one factor of $\bK\times\beps^{*}$. The amplitude is therefore
a pure $M1$, it reads
\begin{equation}
\begin{gathered}
-i\,\mathcal M
= e\,\tilde A\;\chi_f^{\dagger}\,\bS\cdot(\bK\times\beps_\gamma^{\,*})\,\chi_i,\\
\tilde A=\sum_X A_{(X)}\quad[\GeV^{-1}].
\end{gathered}
\label{eq:Mstructure}
\end{equation}
No $E2$ appears at this order. A measured $E2$ part would not fit the minimal
molecular picture.

{\color{black} We now give the covariant form of the amplitude. We follow the
analysis of Ref.~\cite{GamermannBaryon2011} and adapt it to the present
$\tfrac32^-\to\tfrac12^-$ pentaquark (baryon) transition. We write the amplitude with an open
photon index,
\begin{equation}
-i\,\mathcal M=\eps^{*}_\mu(K)\,\mathcal M^{\mu},
\qquad
\mathcal M^{\mu}=\bar u(Q)\,\Gamma^{\mu\nu}\,u_\nu(P),
\label{eq:covamp}
\end{equation}
with $u_\nu(P)$ the Rarita-Schwinger vector-spinor of the $\Pcp$ and $\bar u(Q)$
the Dirac spinor of the $\Pc$. The photon is real and transverse, $K^2=0$ and
$K^\mu\eps^{*}_\mu=0$. The vector-spinor obeys $P^\nu u_\nu=0$ and
$\gamma^\nu u_\nu=0$, so the index $\nu$ reaches the amplitude only through
$g^{\mu\nu}$ and $K^\nu$. Current conservation requires $K_\mu\mathcal M^{\mu}=0$.
These conditions leave two on-shell structures, the magnetic dipole and the
electric quadrupole. In a Jones-Scadron basis \cite{Jones:1972ky}, one finds
\begin{eqnarray}
\Gamma_{M1}^{\mu\nu}&=&\frac{g_{M1}}{2M_i}\,
\varepsilon^{\mu\nu\alpha\beta}P_\alpha K_\beta,
\nonumber\\
\Gamma_{E2}^{\mu\nu}&=&\frac{g_{E2}}{(2M_i)^{2}}\,
\big[(P\cdot K)\,g^{\mu\nu}-P^\mu K^\nu \big]\,\gamma_5\,K\!\!\!/,
\label{eq:M1E2}
\end{eqnarray}
where $g_{M1}$ and $g_{E2}$ set the normalization. Both obey
$K_\mu\Gamma^{\mu\nu}=0$. Both carry the index $\nu$ only through $g^{\mu\nu}$ and
$K^\nu$. The magnetic structure uses the antisymmetric tensor %and needs no$\gamma_5$. 
whereas the electric-quadrupole structure carries $\gamma_5\,K\!\!\!/$ and one more power of momentum.

The two baryons impose which multipoles appear. Both the $\Pcp$ and the $\Pc$ have negative parity, so their relative parity is even. The selection rule then allows $M1$ and $E2$ and forbids $E1$ and $M2$. This is the same content as the $\Delta\to N\gamma$ transition, which also has even relative parity. In Ref.~\cite{GamermannBaryon2011} the two baryons carry opposite parity. The relative parity there is odd, and the leading multipole is the electric dipole carried by the convection current. Here the relative parity is even, the electric dipole is forbidden, and the transition is the magnetic dipole.

All hadronic vertices here are $S$-wave and carry no momentum. The
electric-quadrupole structure of Eq.~\eqref{eq:M1E2} needs the extra power of momentum, so it is not produced and the amplitude is a pure $M1$ mode. In the $\Pcp$ rest frame, with the spin transition operator $\bm S$ and the Coulomb gauge, the magnetic structure of Eq.~\eqref{eq:M1E2} reduces to
$\bm S\cdot(\bK\times\beps^{*})$ of Eq.~\eqref{eq:Mstructure}.}

{\color{black} In this work, we will employ the gauge invariance to constrain the types of diagrams in the amplitudes. The photon couples to the charged meson line, to the charged baryon line, and to the strong $P_cMB$ vertex itself. The last coupling is the contact term of Fig.~\ref{fig:topology}(c). It is the Kroll-Ruderman seagull obtained by gauging the strong vertex through minimal substitution $\partial_\mu\to\partial_\mu-ieQA_\mu$. The hadronic vertices of Eqs.~\eqref{eq:v32V} \textcolor{black}{to} \eqref{eq:v12Vs} are $S$-wave and contain no derivative of the meson field. The minimal substitution then acts only on the meson and baryon kinetic terms and gives the convective (minimal) couplings on the two internal lines. It does not give current at the $P_cMB$ vertex. The contact amplitude of Fig.~\ref{fig:topology}(c) is therefore identically zero,
\begin{equation}
\mathcal M^{\mu}_{(c)}=0 \,.
\label{eq:contactzero}
\end{equation}
As an independent check, the gauge-restoring partner of the convective (minimal coupling) meson current carries the electric structure $\propto\beps_\gamma^{*}$ and not $\bK\times\beps_\gamma^{*}$. After performing loop integration, the only vector left in the $\Pcp$ rest frame is $\bm K$. This structure projects onto $\bm K\cdot\beps_\gamma=0$ and then vanishes. The parity consideration also gives the same result. The electric structure $\propto\beps_\gamma^{*}$ is the electric dipole
$\bm S\cdot\beps_\gamma^{*}$, and the even relative parity of the two baryons forbids it. The convective couplings on the
meson and baryon lines drop in the same way, as shown in
App.~\ref{app:gauge-conv}. Only the transverse magnetic vertices remain. Each one is proportional to $\bK\times\beps^{*}$, so every triangle is gauge invariant on its own, and the loops add directly.}

\textcolor{black}{In this work, we do not introduce an additional phenomenological form-factor at the $P_cMB$ vertices. The present calculation is based on the chiral-unitary description of the two poles. The couplings $g_i$ are the pole residues of the same meson-baryon amplitude that generates the molecular states
\cite{XNO2019}. The short-distance part is therefore already fixed by the regularization of the loop function used in the coupled-channel problem.  Similar radiative transitions of dynamically generated states are commonly evaluated by coupling the photon to the hadronic constituents in the meson-baryon loops
\cite{GamermannBaryon2011,Jido:2007sm,Doring:2007rz}. An independent vertex form-factor would introduce an additional short-distance model dependence. It would also require the corresponding contact current in order to preserve the Ward-Takahashi identity
\cite{Haberzettl:1997jg,Davidson:2001rk,Haberzettl:2021vmd}. At the order considered here the transition is the magnetic $M1$ amplitude. \textcolor{black}{The contact current that a derivative vertex would require is absent for our $S$-wave
vertices. This is shown above through Eq.~\eqref{eq:contactzero} and in App.~\ref{app:gauge-conv}.}} In the latter, however, we will include the form-factor effect in the loop amplitudes in this work in order to compare the results with those in the literature.

%---------------------------------------------------------------------
\subsection{Radiative decay amplitudes}
\label{sec:amps}
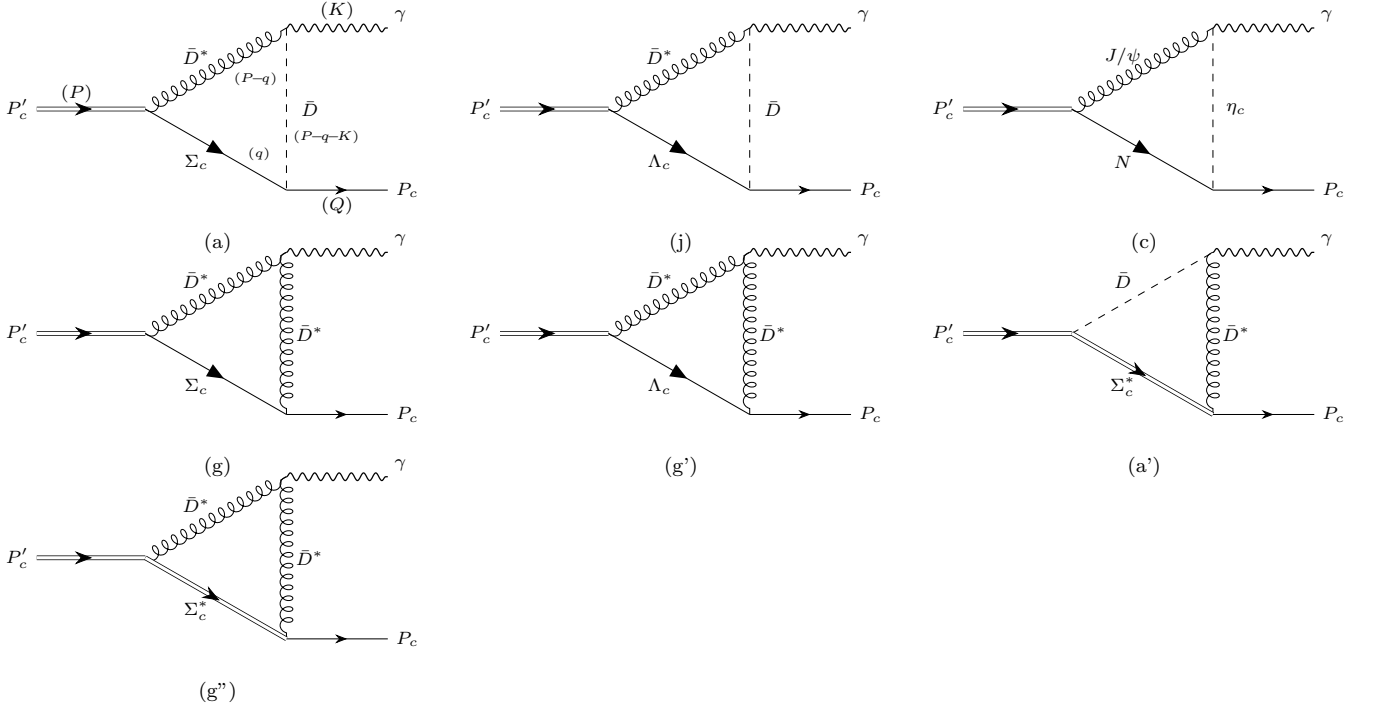
\begin{figure*}[tb]
\centering
\resizebox{\textwidth}{!}{%
\begin{tikzpicture}[every node/.style={font=\small}]
\begin{scope}\pcf{a}\triMbb{vector}{pseudoscalar}\lUL{\(\bar D^{*}\)}\lV{\(\bar D\)}\lLL{\(\Sigma_c\)}
  \node[font=\scriptsize] at (-0.95,0.22){\((P)\)};
  \node[font=\scriptsize] at ($(BR)!0.5!(F)+(0,-0.22)$){\((Q)\)};
  \node[font=\scriptsize] at ($(TR)!0.5!(G)+(0,0.24)$){\((K)\)};
  \node[font=\scriptsize,black] at ($(L)!0.55!(TR)+(0.45,-0.2)$){\(\scriptstyle (P\!-\!q)\)};
  \node[font=\scriptsize,black] at (2.5,-0.39){\(\scriptstyle (P\!-\!q\!-\!K)\)};
  \node[font=\scriptsize,black] at ($(L)!0.55!(BR)+(0.5,0.0)$){\(\scriptstyle (q)\)};\end{scope}
\begin{scope}[xshift=6.4cm]\pcf{j}\triMbb{vector}{pseudoscalar}\lUL{\(\bar D^{*}\)}\lV{\(\bar D\)}\lLL{\(\Lambda_c\)}\end{scope}
\begin{scope}[xshift=12.8cm]\pcf{c}\triMbb{vector}{pseudoscalar}\lUL{\(J/\psi\)}\lV{\(\eta_c\)}\lLL{\(N\)}\end{scope}
\begin{scope}[yshift=-3.1cm]\pcf{g}\triMbb{vector}{vector}\lUL{\(\bar D^{*}\)}\lV{\(\bar D^{*}\)}\lLL{\(\Sigma_c\)}\end{scope}
\begin{scope}[xshift=6.4cm,yshift=-3.1cm]\pcf{g'}\triMbb{vector}{vector}\lUL{\(\bar D^{*}\)}\lV{\(\bar D^{*}\)}\lLL{\(\Lambda_c\)}\end{scope}
\begin{scope}[xshift=12.8cm,yshift=-3.1cm]\pcf{a'}\triMbB{pseudoscalar}{vector}\lUL{\(\bar D\)}\lV{\(\bar D^{*}\)}\lLL{\(\Sigma_c^{*}\)}\end{scope}
\begin{scope}[yshift=-6.2cm]\pcf{g''}\triMbB{vector}{vector}\lUL{\(\bar D^{*}\)}\lV{\(\bar D^{*}\)}\lLL{\(\Sigma_c^{*}\)}\end{scope}
\end{tikzpicture}}
\caption{The seven class-M (meson-line photon) triangle loops. The $P_c(4457)$
(spin-$\tfrac32$, double line) enters at the left apex, the photon leaves the
upper-right vertex, and the $P_c(4312)$ (spin-$\tfrac12$, plain) leaves the
lower-right vertex. Vector mesons are curly, pseudoscalars dashed,
spin-$\tfrac32$ baryons doubled, spin-$\tfrac12$ baryons plain. Four-momenta
are shown on (a) with $P=K+Q$ and the loop momentum $q$ on the spectator.}
\label{fig:triangles}
\end{figure*}

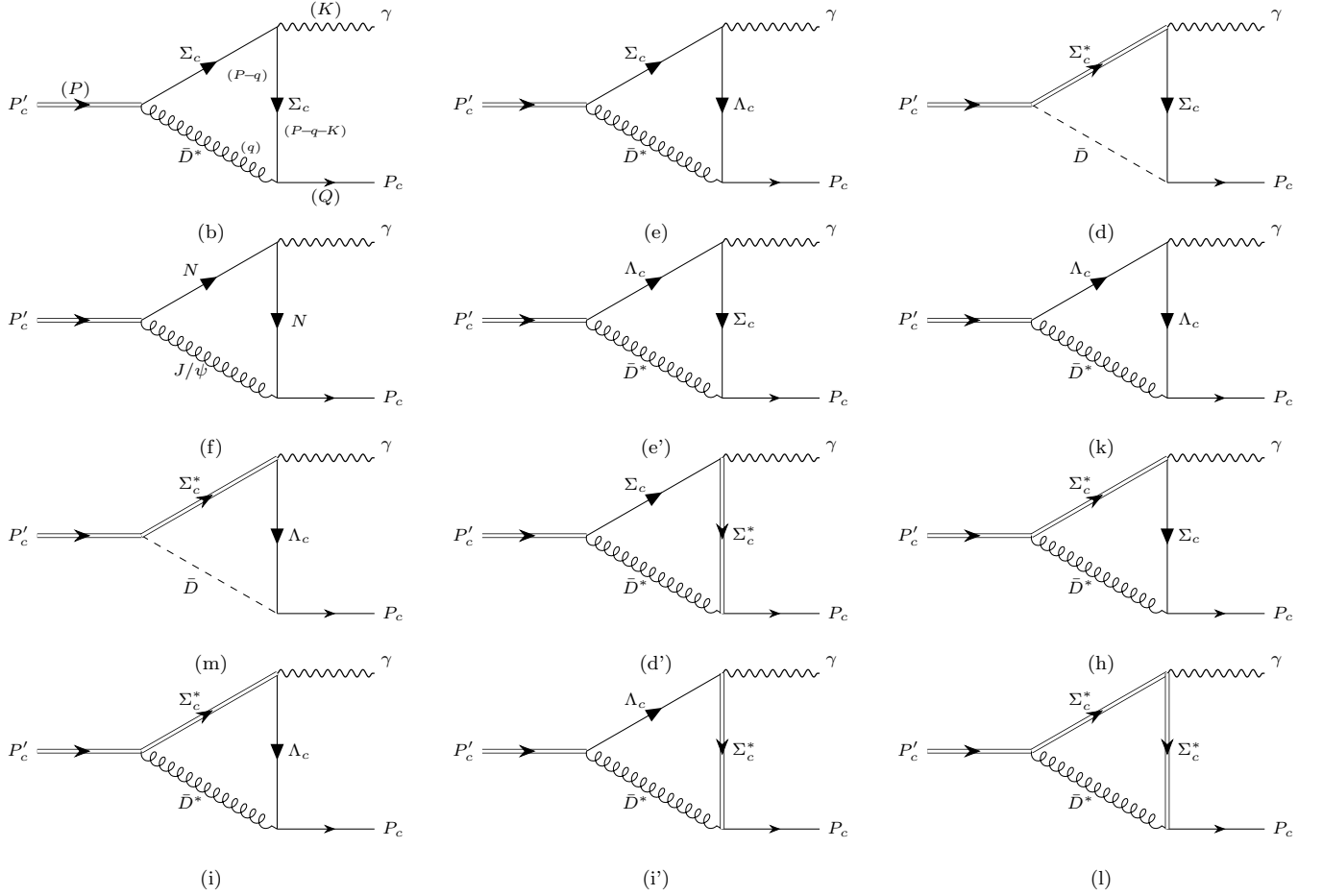
\begin{figure*}[tb]
\centering
\resizebox{\textwidth}{!}{%
\begin{tikzpicture}[every node/.style={font=\small}]
\begin{scope}\pcf{b}\triBbb{vector}\lUL{\(\Sigma_c\)}\lV{\(\Sigma_c\)}\lLL{\(\bar D^{*}\)}
\node[font=\scriptsize] at (-0.95,0.22){\((P)\)};
  \node[font=\scriptsize] at ($(BR)!0.5!(F)+(0,-0.22)$){\((Q)\)};
  \node[font=\scriptsize] at ($(TR)!0.5!(G)+(0,0.24)$){\((K)\)};
  \node[font=\scriptsize,black] at ($(L)!0.55!(TR)+(0.45,-0.2)$){\(\scriptstyle (P\!-\!q)\)};
  \node[font=\scriptsize,black] at (2.5,-0.39){\(\scriptstyle (P\!-\!q\!-\!K)\)};
  \node[font=\scriptsize,black] at ($(L)!0.55!(BR)+(0.5,0.0)$){\(\scriptstyle (q)\)};\end{scope}
\begin{scope}[xshift=6.4cm]\pcf{e}\triBbb{vector}\lUL{\(\Sigma_c\)}\lV{\(\Lambda_c\)}\lLL{\(\bar D^{*}\)}\end{scope}
\begin{scope}[xshift=12.8cm]\pcf{d}\triBBb{pseudoscalar}\lUL{\(\Sigma_c^{*}\)}\lV{\(\Sigma_c\)}\lLL{\(\bar D\)}\end{scope}
\begin{scope}[yshift=-3.1cm]\pcf{f}\triBbb{vector}\lUL{\(N\)}\lV{\(N\)}\lLL{\(J/\psi\)}\end{scope}
\begin{scope}[xshift=6.4cm,yshift=-3.1cm]\pcf{e'}\triBbb{vector}\lUL{\(\Lambda_c\)}\lV{\(\Sigma_c\)}\lLL{\(\bar D^{*}\)}\end{scope}
\begin{scope}[xshift=12.8cm,yshift=-3.1cm]\pcf{k}\triBbb{vector}\lUL{\(\Lambda_c\)}\lV{\(\Lambda_c\)}\lLL{\(\bar D^{*}\)}\end{scope}
\begin{scope}[yshift=-6.2cm]\pcf{m}\triBBb{pseudoscalar}\lUL{\(\Sigma_c^{*}\)}\lV{\(\Lambda_c\)}\lLL{\(\bar D\)}\end{scope}
\begin{scope}[xshift=6.4cm,yshift=-6.2cm]\pcf{d'}\triBbB{vector}\lUL{\(\Sigma_c\)}\lV{\(\Sigma_c^{*}\)}\lLL{\(\bar D^{*}\)}\end{scope}
\begin{scope}[xshift=12.8cm,yshift=-6.2cm]\pcf{h}\triBBb{vector}\lUL{\(\Sigma_c^{*}\)}\lV{\(\Sigma_c\)}\lLL{\(\bar D^{*}\)}\end{scope}
\begin{scope}[yshift=-9.3cm]\pcf{i}\triBBb{vector}\lUL{\(\Sigma_c^{*}\)}\lV{\(\Lambda_c\)}\lLL{\(\bar D^{*}\)}\end{scope}
\begin{scope}[xshift=6.4cm,yshift=-9.3cm]\pcf{i'}\triBbB{vector}\lUL{\(\Lambda_c\)}\lV{\(\Sigma_c^{*}\)}\lLL{\(\bar D^{*}\)}\end{scope}
\begin{scope}[xshift=12.8cm,yshift=-9.3cm]\pcf{l}\triBBB{vector}\lUL{\(\Sigma_c^{*}\)}\lV{\(\Sigma_c^{*}\)}\lLL{\(\bar D^{*}\)}\end{scope}
\end{tikzpicture}}
\caption{The twelve class-B (baryon-line photon) triangle loops, same
conventions as Fig.~\ref{fig:triangles}. The photon is emitted from the baryon
line. The spectator meson is the lower-left edge. Spin-$\tfrac32$ baryons
($\Scs$) are drawn as double lines.}
\label{fig:trianglesB}
\end{figure*}

Gauge invariance requires for the sum over the full channel content of both poles
\cite{GamermannMeson2007,GamermannBaryon2011,NagahiroRocaHosakaOset2009}. A triangle loop contributes when the
$\Pcp$ couples to a channel $M_1B_1$, a photon converts $M_1B_1\to M_2B_2$ by radiating off the meson or the baryon, and the $\Pc$ couples to $M_2B_2$. We go through every meson-baryon pair of the two states. This gives nineteen triangle loops. We sort them into two classes. On the one hand, in class M the photon couples to a meson line and the baryon is the spectator. On the other hand, in class B the photon couples to a baryon line and the meson is the spectator. Figure~\ref{fig:topology} shows the two triangle topologies. \textcolor{black}{It also shows the contact term of
Fig.~\ref{fig:topology}(c), which vanishes for our $S$-wave vertices and is not counted among the nineteen loops.} Figures~\ref{fig:triangles} and \ref{fig:trianglesB} draw all
nineteen. Tab.~\ref{tab:loops} lists them. In addition, the elastic $J/\psi$ loop vanishes by $C$ parity.
%%%%%%%%%%%%%%%%%%%%%%%%%%%%%%%%%%%%%%%%%%%%%%%%%%%%%%%%%%%%%%%%%%%%%%%%%%%%%%%%%%%%%%%%%%%%%%%%%%%%%%%%%%%%%%%%%%%%%%%%%%%%%%%%%%%%%%%%%%%%%%%%
\begin{table*}[tb]
\centering
\renewcommand{\arraystretch}{1.3}
\setlength{\tabcolsep}{10pt}
\begin{tabular}{c l c c c}
\toprule
$X$ & process & $(g_ig_f)_X$ & photon vertex & $\mathfrak{s}_X$\\
\midrule
\multicolumn{5}{l}{\textit{Class M\,: photon emitted from the meson line}}\\
\cmidrule(lr){1-5}
(a)    & $\Dbs\Sc\to[\Dbs\!\to\!\Db\gamma]\to\Db\Sc$            & $g_1\,g_2$                          & $\lambda_{0},\lambda_-$ & $+1$\\
(g)    & $\Dbs\Sc\to[\Dbs\ \mathrm{el.}]\to\Dbs\Sc$            & $g_1\,g_3$                          & $\lambda_g$             & $-1$\\
(j)    & $\Dbs\Lc\to[\Dbs\!\to\!\Db\gamma]\to\Db\Lc$           & $g^{i}_{\Dbs\Lc}\,g^{f}_{\Db\Lc}$   & $\lambda_0$             & $+1$\\
(g$'$) & $\Dbs\Lc\to[\Dbs\ \mathrm{el.}]\to\Dbs\Lc$            & $g^{i}_{\Dbs\Lc}\,g^{f}_{\Dbs\Lc}$  & $\lambda_g$             & $-1$\\
(c)    & $J/\psi\,N\to[\psi\!\to\!\eta_c\gamma]\to\eta_c N$     & $g^{i}_{\psi}\,g^{f}_{\eta}$        & $\lambda_\psi$          & $+1$\\
(a$'$) & $\Db\Scs\to[\Db\!\to\!\Dbs\gamma]\to\Dbs\Scs$         & $g^{i}_{\Db\Scs}\,g^{f}_{\Dbs\Scs}$ & $\lambda_{0},\lambda_-$ & $+\tfrac12^{\dagger}$\\
(g$''$)& $\Dbs\Scs\to[\Dbs\ \mathrm{el.}]\to\Dbs\Scs$          & $g^{i}_{\Dbs\Scs}\,g^{f}_{\Dbs\Scs}$& $\lambda_g$             & $-\tfrac13^{\dagger}$\\
\midrule
\multicolumn{5}{l}{\textit{Class B\,: photon emitted from the baryon line}}\\
\cmidrule(lr){1-5}
(b)    & $\Dbs\Sc\to[\Sc\ \mathrm{el.}]\to\Dbs\Sc$             & $g_1\,g_3$                          & $\tilde\mu_{\Sc}$       & $+2$\\
(e)    & $\Dbs\Sc\to[\Sc^{+}\!\to\!\Lc^{+}\gamma]\to\Dbs\Lc$   & $g_1\,g^{f}_{\Dbs\Lc}$              & $\tilde\mu_{\Sc\Lc}$    & $+2$\\
(d$'$) & $\Dbs\Sc\to[\Sc\!\to\!\Scs\gamma]\to\Dbs\Scs$         & $g_1\,g^{f}_{\Dbs\Scs}$             & $\tilde\mu_{3}$         & $+\tfrac13$\\
(e$'$) & $\Dbs\Lc\to[\Lc^{+}\!\to\!\Sc^{+}\gamma]\to\Dbs\Sc$   & $g^{i}_{\Dbs\Lc}\,g_3$              & $\tilde\mu_{\Sc\Lc}$    & $+2$\\
(k)    & $\Dbs\Lc\to[\Lc\ \mathrm{el.}]\to\Dbs\Lc$             & $g^{i}_{\Dbs\Lc}\,g^{f}_{\Dbs\Lc}$  & $\tilde\mu_{\Lc}$       & $+2$\\
(i$'$) & $\Dbs\Lc\to[\Lc\!\to\!\Scs\gamma]\to\Dbs\Scs$         & $g^{i}_{\Dbs\Lc}\,g^{f}_{\Dbs\Scs}$ & $\tilde\mu_{\Scs\Lc}$   & $+1^{\dagger}$\\
(f)    & $J/\psi\,N\to[p\ \mathrm{el.}]\to J/\psi\,N$           & $g^{i}_{\psi}\,g^{f}_{\psi}$        & $\tilde\mu_{p}$         & $+2$\\
(d)    & $\Db\Scs\to[\Scs\!\to\!\Sc\gamma]\to\Db\Sc$           & $g^{i}_{\Db\Scs}\,g_2$              & $\tilde\mu_{3}$         & $+1$\\
(m)    & $\Db\Scs\to[\Scs\!\to\!\Lc\gamma]\to\Db\Lc$           & $g^{i}_{\Db\Scs}\,g^{f}_{\Db\Lc}$   & $\tilde\mu_{\Scs\Lc}$   & $+1$\\
(h)    & $\Dbs\Scs\to[\Scs\!\to\!\Sc\gamma]\to\Dbs\Sc$         & $g^{i}_{\Dbs\Scs}\,g_3$             & $\tilde\mu_{3}$         & $+1^{\dagger}$\\
(i)    & $\Dbs\Scs\to[\Scs\!\to\!\Lc\gamma]\to\Dbs\Lc$         & $g^{i}_{\Dbs\Scs}\,g^{f}_{\Dbs\Lc}$ & $\tilde\mu_{\Scs\Lc}$   & $+1^{\dagger}$\\
(l)    & $\Dbs\Scs\to[\Scs\ \mathrm{el.}]\to\Dbs\Scs$          & $g^{i}_{\Dbs\Scs}\,g^{f}_{\Dbs\Scs}$& $\tilde\mu_{\Scs}$      & $+1^{\dagger}$\\
\bottomrule
\end{tabular}
\caption{The nineteen $M1$ triangle loops that build the transition
$P_c(4457)\to P_c(4312)\gamma$. Each loop carries the label $X$ used in
Figs.~\ref{fig:triangles} and \ref{fig:trianglesB} and in
Tab.~\ref{tab:explicit}. The process column reads as
[\,$P_c(4457)$ channel\,]\,$\to$\,[\,radiating transition\,]\,$\to$\,[\,$P_c(4312)$ channel\,],
where ``el.'' denotes a diagonal magnetic-moment vertex. The loops are grouped by
the line that emits the photon. In class M the photon leaves the meson line and
the baryon is the spectator. In class B the photon leaves the baryon line and the
meson is the spectator. Within each class the loops are ordered by the initial
channel. Here $(g_ig_f)_X$ is the product of the two molecular residues of
Tab.~\ref{tab:residues}, with $i$ and $f$ the initial and final couplings.
The photon coupling is $\lambda$ for a heavy-vector $M1$ vertex of
Eq.~\eqref{eq:lamdef} ($\lambda_0$ and $\lambda_-$ for the neutral and charged
charge states, $\lambda_\psi$ for $J/\psi\to\eta_c\gamma$, $\lambda_g$ for the
elastic vector moment) and $\tilde\mu$ for a baryon magnetic or transition
moment of App.~\ref{app:em}. The spin factor $\mathfrak{s}_X$ multiplies the
universal structure $\chi_f^{\dagger}\,\bm S\cdot(\bm K\times\beps^{*})\,\chi_i$
and is derived in App.~\ref{app:spin}. The six entries marked with a dagger
carry both a $\Scs$ and a $\Dbs$ in the triangle, a $\tfrac32\otimes1$
recoupling. Their spin factors are estimated at leading order and are shown in
App.~\ref{app:spin} to be numerically immaterial, a simultaneous change of
all six shifting the width by $2.7\%$. The
elastic $J/\psi$ loop vanishes by $C$ parity and is absent.}
\label{tab:loops}
\end{table*}
%%%%%%%%%%%%%%%%%%%%%%%%%%%%%%%%%%%%%%%%%%%%%%%%%%%%%%%%%%%%%%%%%%%%%%%%%%%%%%%%%%%%%%%%%%%%%%%%%%%%%%%%%%%%%%%%%%%%%%%%%%%%%%%%%%%%%%%%%%%%%%%%
%%%%%%%%%%%%%%%%%%%%%%%%%%%%%%%%%%%%%%%%%%%%%%%%%%%%%%%%%%%%%%%%%%%%%%%%%%%%%%%%%%%%%%%%%%%%%%%%%%%%%%%%%%%%%%%%%%%%%%%%%%%%%%%%%%%%%%%%%%%%%%%%
%%%%%%%%%%%%%%%%%%%%%%%%%%%%%%%%%%%%%%%%%%%%%%%%%%%%%%%%%%%%%%%%%%%%%%%%%%%%%%%%%%%%%%%%%%%%%%%%%%%%%%%%%%%%%%%%%%%%%%%%%%%%%%%%%%%%%%%%%%%%%%%%
\paragraph{Numerators.}
The numerator that multiplies the three propagators in Eq.~\eqref{eq:tridef} follows from the molecular and photon vertices. We write the universal $M1$
structure as $\mathcal T\equiv\chi_f^{\dagger}\bm S\cdot(\bm K\times\beps^{*})\chi_i$.
For class M (photon on the meson line) the numerator is linear in the radiating
meson momentum $(P-q)^\nu$, for example for loop (a),
\begin{equation}
\mathcal N^{(M)}_{a}(q)=2M_{\Sc}\,\lambda\;
S^{\rho}\,\varepsilon_{\rho\nu\alpha\beta}\,(P-q)^{\nu}\,\eps^{*\alpha}\,K^{\beta}.
\label{eq:NMa}
\end{equation}
For class B (photon on the baryon line) the two baryon propagators give the
constants $2M_{B_1}$ and $2M_{B_2}$, the magnetic vertex carries only the
external $\bm K,\beps^{*}$, and the numerator is independent of $q$, for example
for loop (b),
\begin{equation}
\begin{aligned}
\mathcal N^{(B)}_{b}&=(2M_{\Sc})^{2}\,\tilde\mu_{\Sc}\;
\chi_f^{\dagger}\,\bm\sigma\cdot\big[\bm\sigma\!\cdot\!(\bm K\times\beps^{*})\big]\,\bm S\,\chi_i\\
&=2\,(2M_{\Sc})^{2}\,\tilde\mu_{\Sc}\,\mathcal T,
\end{aligned}
\label{eq:NBb}
\end{equation}
where $\bm \sigma\cdot(\bm\sigma\!\cdot\!\bm A)\,\bm S=2\,\bm S\!\cdot\!\bm A$ gives the
spin factor $2$, see App.~\ref{app:spin}.

\paragraph{A generic form of the triangle loop amplitudes.}
{\textcolor{black}{With the loop momentum $q$ and the momentum routing shown in the figures, the three denominators of the triangle loop amplitudes are}}
\begin{eqnarray}
D_1&=&(P-q)^2-M_1^2+i\eps,
\nonumber\\
D_2&=&(P-q-K)^2-M_2^2+i\eps,
\nonumber\\
D_3&=&q^2-M_s^2+i\eps.
\label{eq:Ddef}   
\end{eqnarray}
The contribution of loop $X$ to the reduced amplitude $\tilde A$ of
Eq.~\eqref{eq:Mstructure} is written by
\begin{equation}
A_{(X)}\,\mathcal T
=i\,(g_ig_f)_X\!\sum_{\alpha}\,\mathcal C_\alpha^{2}\!
\int\!\frac{d^4q}{(2\pi)^4}\,
\frac{\mathcal N^{(\kappa)}_{X,\alpha}(q)}{D_{1,\alpha}\,D_{2,\alpha}\,D_{3,\alpha}}\,.
\label{eq:Apre}
\end{equation}
where $\kappa=\{M,B\}$, $\alpha$ runs over the charge states of the internal channel, $\mathcal C_\alpha^{2}$
is the isospin weight ($\tfrac13$ neutral, $\tfrac23$ charged, $1$ single state), and $(g_ig_f)_X$ is the product of the two hadronic coupling residues including
the Cartesian factors $1/\sqrt3$, $1/\sqrt2$ and the $I=1\to0$ recoupling
$1/\sqrt3$. The numerator is Eq.~\eqref{eq:NMa} for class M, with general factor
$\mathfrak s_X(2M_{s,\alpha})\lambda_c$, and is $q$-independent
$\mathfrak s_X(2M_{1,\alpha})(2M_{2,\alpha})\tilde\mu_\alpha\,\mathcal T$ for class B. As two
explicit cases, loop (a) and loop (b) read
\begin{align}
A_{(a)}&=g_1 g_2\,\frac{M_i}{16\pi^{2}}
\sum_{\alpha} C_\alpha^{2}\,\lambda_\alpha\,(2M_{\Sc^\alpha})\,I_a^{\alpha},
\nonumber\\
I_a^{\alpha}&=\int_0^1\!\!dx\!\int_0^{1-x}\!\!dy\;\frac{y}{s_a^{\alpha}(x,y)}\,,
\label{eq:Aa}\\
A_{(b)}&=2\,g_1\,\frac{g_3}{\sqrt3}\,\frac{1}{16\pi^{2}}
\sum_{\alpha} C_\alpha^{2}\,\tilde\mu_{\Sc^\alpha}\,(2M_{\Sc^\alpha})^{2}\,I_b^{\alpha},
\nonumber\\
I_b^{\alpha}&=\int_0^1\!\!dx\!\int_0^{1-x}\!\!dy\;\frac{1}{s_b^{\alpha}(x,y)}\,.
\label{eq:Ab}
\end{align}
The remaining seventeen amplitudes have the same form. Each one is fixed by its row in Tab.~\ref{tab:explicit}. The additionally dominant loop is (e). It radiates the transition moment $\Sc^{+}\to\Lc^{+}\gamma$ and reads
\begin{equation}
A_{(e)}=2\,g_1\,\frac{g_{\Dbs\Lc}^{f}}{\sqrt3}\frac{1}{\sqrt3}\,
\frac{1}{16\pi^2}\,\tilde\mu_{\Sc\Lc}\,(2M_{\Sc^+})(2M_{\Lc})\,I_e.
\label{eq:Ae}
\end{equation}
This amplitude is unexpectedly large given the small value of the residue $g_{\Dbs\Lc}^{f}=0.03+0.25i$ because the
pair $\bar D^{*0}\Sc^{+}$ is on the threshold and the exit channel $\Dbs\Lc$ is open, so $I_e$ has a large imaginary part. Two contributions vanish at leading order and are not in Tab.~\ref{tab:loops}. The charge-convection term on a
meson line vanishes in the rest-frame Coulomb gauge. The elastic $J/\psi$ loop vanishes by $C$ parity.

\paragraph{Feynman parametrization and the final analytic form of the loop amplitudes.}
Combine the three denominators with the Feynman parameterization. Assign the parameter $y$ to the spectator line $D_3$, $x$ to the post-photon line $D_2$ and $1-x-y$ to the entering line $D_1$,
\begin{equation}
\frac{1}{D_1D_2D_3}=\int_0^1\!\!dx\!\int_0^{1-x}\!\!dy\;
\frac{2}{\big[(1-x-y)D_1+xD_2+yD_3\big]^{3}}.
\label{eq:feyn}
\end{equation}
Completing the square fixes the shift $\bar q=(1-y)P-xK$. After shifting the loop momentum, the
denominator depends on $(x,y)$ only through the quadratic
\begin{equation}
s_{X,\alpha}(x,y)=-M_i^{2}\,y^{2}+b_{X,\alpha}(x)\,y+c_{X,\alpha}(x),
\label{eq:smaster}
\end{equation}
with
\begin{eqnarray}
b_{X,\alpha}(x)&=&M_i^{2}-2x\,M_iE_\gamma-M_{s,\alpha}^{2}+M_{1,\alpha}^{2},
\nonumber\\
c_{X,\alpha}(x)&=&-x\,M_{2,\alpha}^{2}-(1-x)\,M_{1,\alpha}^{2}+i\eps\,.
\label{eq:bcdef}
\end{eqnarray}
The Wick-rotated scalar integral is
$i\int d^4q/(2\pi)^4\,[(q-\bar q)^2-\Delta]^{-3}=-1/(32\pi^2 s)$ with
$\Delta=-s$, and the factor $2$ from Feynman trick gives $1/(16\pi^2)$ of the amplitudes. The numerator class fixes the weight factor $w_\kappa$. For class M the meson
momentum becomes $(P-q)^\nu=yP^\nu+xK^\nu-\ell^\nu$ after the shift. The
$\ell$-odd part integrates to zero, the $xK^\nu$ part vanishes against the
Levi-Civita $K^\beta$, and only $yP^\nu$ survives, giving
$S^\rho\eps_{\rho\nu\alpha\beta}P^\nu\eps^{*\alpha}K^\beta=M_i\,\mathcal T$ and
the weight $y$. For class B the numerator carries no $q$ and the weight is $1$.
Each amplitude then reduces to the two-parameter integral
\begin{widetext}
\begin{equation}
A_{(X)}=\mathfrak{s}_X\,(g_ig_f)_X\,\frac{(M_i)^{\,\delta_{\kappa M}}}{16\pi^{2}}
\sum_{\alpha}\mathcal C_\alpha^{2}\;\Xi^{(\kappa)}_{X,\alpha}
\int_0^1\!\!dx\int_0^{1-x}\!\!dy\;\frac{w_\kappa}{s_{X,\alpha}(x,y)},
\label{eq:Apost}
\end{equation}
%\end{widetext}
with $w_M=y$, $w_B=1$, the prefactor $M_i$ present only for class M, and the
mass and EM factor
\begin{equation}
\Xi^{(M)}_{X,\alpha}=\lambda_{\alpha}\,(2M_{s,\alpha}),\qquad
\Xi^{(B)}_{X,\alpha}=\tilde\mu_{\alpha}\,(2M_{1,\alpha})(2M_{2,\alpha}).
\label{eq:Xidef}
\end{equation}
The inner $y$-integral is done in closed form. With the roots
$r_{1,2}=(-b_{X,\alpha}\pm\sqrt{b_{X,\alpha}^{2}-4ac_{X,\alpha}})/2a$ and $a=-M_i^2$,
%\begin{widetext}
\begin{align}
\int_0^{1-x}\!\frac{y\,dy}{s}
&=\frac{1}{a(r_1-r_2)}\Big[r_1\ln\tfrac{(1-x)-r_1}{-r_1}
-r_2\ln\tfrac{(1-x)-r_2}{-r_2}\Big]
\quad(\text{class M}),
\label{eq:yint2}
\\
\int_0^{1-x}\!\frac{dy}{s}
&=\frac{1}{a(r_1-r_2)}\Big[\ln\tfrac{(1-x)-r_1}{-r_1}
-\ln\tfrac{(1-x)-r_2}{-r_2}\Big]
\quad(\text{class B}).
\label{eq:yint1}
\end{align}
\end{widetext}
The remaining $x$ integral is done by quadrature on the real and imaginary parts. When all internal channels are closed, $s<0$ on the triangle and the amplitude is real. When a channel is open, the $+i\eps$ in $c_{X,\alpha}$ selects the branch of the logarithm and gives the absorptive part of the open $J/\psi N$,
$\eta_c N$ and $\Dbs\Lc$ states. The full closed form for all nineteen amplitudes and a numerical test value are given in App.~\ref{app:yint}.

The complete specification of the nineteen amplitudes is Tab.~\ref{tab:explicit}. Inserting any row into Eq.~\eqref{eq:Apost} gives that loop in its final form.
% Table IV. Requires \usepackage{multirow} in the preamble.
\begin{table*}[tb]
\centering
\renewcommand{\arraystretch}{1}
\setlength{\tabcolsep}{10pt}
\footnotesize
\begin{tabular}{c c c c c c c}
\toprule
$X$ & $(g_ig_f)_X$ & $n_X$ & $\mathfrak{s}_X$ & $\alpha$ &
$(\mathcal{C}_\alpha^{2};\,\mathrm{EM}_\alpha)$ & $(M_1,M_2,M_s)\ [\mathrm{GeV}]$\\
\midrule
\multicolumn{7}{l}{\textit{Class M\,: photon on the meson line, $w_M=y$}}\\
\cmidrule(lr){1-7}
\multirow{2}{*}{(a)} & \multirow{2}{*}{$g_1 g_2$} & \multirow{2}{*}{$1$} & \multirow{2}{*}{$+1$}
 & $n$ & $(1/3;\,\lambda_0)$ & $(2.00685,\,1.86484,\,2.45290)$\\
 & & & & $c$ & $(2/3;\,\lambda_-)$ & $(2.01026,\,1.86966,\,2.45397)$\\
\addlinespace[1pt]
\multirow{2}{*}{(g)} & \multirow{2}{*}{$g_1 g_3$} & \multirow{2}{*}{$1/\sqrt3$} & \multirow{2}{*}{$-1$}
 & $n$ & $(1/3;\,\lambda_g^{0})$ & $(2.00685,\,2.00685,\,2.45290)$\\
 & & & & $c$ & $(2/3;\,\lambda_g^{-})$ & $(2.01026,\,2.01026,\,2.45397)$\\
\addlinespace[1pt]
\multirow{2}{*}{(j)} & \multirow{2}{*}{$g^{i}_{\Dbs\Lc}\,g^{f}_{\Db\Lc}$} & \multirow{2}{*}{$1$} & \multirow{2}{*}{$+1$}
 & $n$ & $(1/3;\,\lambda_0)$ & $(2.00685,\,1.86484,\,2.28646)$\\
 & & & & $c$ & $(2/3;\,\lambda_-)$ & $(2.01026,\,1.86966,\,2.28646)$\\
\addlinespace[1pt]
\multirow{2}{*}{(g$'$)} & \multirow{2}{*}{$g^{i}_{\Dbs\Lc}\,g^{f}_{\Dbs\Lc}$} & \multirow{2}{*}{$1/\sqrt3$} & \multirow{2}{*}{$-1$}
 & $n$ & $(1/3;\,\lambda_g^{0})$ & $(2.00685,\,2.00685,\,2.28646)$\\
 & & & & $c$ & $(2/3;\,\lambda_g^{-})$ & $(2.01026,\,2.01026,\,2.28646)$\\
\addlinespace[1pt]
(c) & $g^{i}_{\psi}\,g^{f}_{\eta}$ & $1$ & $+1$ & $-$ & $(1;\,\lambda_\psi)$ & $(3.09690,\,2.98390,\,0.93827)$\\
\addlinespace[1pt]
\multirow{2}{*}{(a$'$)} & \multirow{2}{*}{$g^{i}_{\Db\Scs}\,g^{f}_{\Dbs\Scs}$} & \multirow{2}{*}{$1$} & \multirow{2}{*}{$+1/2^{\dagger}$}
 & $n$ & $(1/3;\,\lambda_0)$ & $(1.86484,\,2.00685,\,2.51750)$\\
 & & & & $c$ & $(2/3;\,\lambda_-)$ & $(1.86966,\,2.01026,\,2.51841)$\\
\addlinespace[1pt]
\multirow{2}{*}{(g$''$)} & \multirow{2}{*}{$g^{i}_{\Dbs\Scs}\,g^{f}_{\Dbs\Scs}$} & \multirow{2}{*}{$1/\sqrt2$} & \multirow{2}{*}{$-1/3^{\dagger}$}
 & $n$ & $(1/3;\,\lambda_g^{0})$ & $(2.00685,\,2.00685,\,2.51750)$\\
 & & & & $c$ & $(2/3;\,\lambda_g^{-})$ & $(2.01026,\,2.01026,\,2.51841)$\\
\midrule
\multicolumn{7}{l}{\textit{Class B\,: photon on the baryon line, $w_B=1$}}\\
\cmidrule(lr){1-7}
\multirow{2}{*}{(b)} & \multirow{2}{*}{$g_1 g_3$} & \multirow{2}{*}{$1/\sqrt3$} & \multirow{2}{*}{$+2$}
 & $n$ & $(1/3;\,\tilde\mu_{\Sc^{+}})$ & $(2.45290,\,2.45290,\,2.00685)$\\
 & & & & $c$ & $(2/3;\,\tilde\mu_{\Sc^{++}})$ & $(2.45397,\,2.45397,\,2.01026)$\\
\addlinespace[1pt]
(e) & $g_1\,g^{f}_{\Dbs\Lc}$ & $1/3$ & $+2$ & $-$ & $(1;\,\tilde\mu_{\Sc\Lc})$ & $(2.45290,\,2.28646,\,2.00685)$\\
\addlinespace[1pt]
\multirow{2}{*}{(d$'$)} & \multirow{2}{*}{$g_1\,g^{f}_{\Dbs\Scs}$} & \multirow{2}{*}{$1/\sqrt2$} & \multirow{2}{*}{$+1/3$}
 & $n$ & $(1/3;\,\tilde\mu_3^{+})$ & $(2.45290,\,2.51750,\,2.00685)$\\
 & & & & $c$ & $(2/3;\,\tilde\mu_3^{++})$ & $(2.45397,\,2.51841,\,2.01026)$\\
\addlinespace[1pt]
(e$'$) & $g^{i}_{\Dbs\Lc}\,g_3$ & $1/3$ & $+2$ & $-$ & $(1;\,\tilde\mu_{\Sc\Lc})$ & $(2.28646,\,2.45290,\,2.00685)$\\
\addlinespace[1pt]
(k) & $g^{i}_{\Dbs\Lc}\,g^{f}_{\Dbs\Lc}$ & $1/\sqrt3$ & $+2$ & $-$ & $(1;\,\tilde\mu_{\Lc})$ & $(2.28646,\,2.28646,\,2.00685)$\\
\addlinespace[1pt]
(i$'$) & $g^{i}_{\Dbs\Lc}\,g^{f}_{\Dbs\Scs}$ & $1/\sqrt6$ & $+1^{\dagger}$ & $-$ & $(1;\,\tilde\mu_{\Scs\Lc})$ & $(2.28646,\,2.51750,\,2.00685)$\\
\addlinespace[1pt]
(f) & $g^{i}_{\psi}\,g^{f}_{\psi}$ & $1/\sqrt3$ & $+2$ & $-$ & $(1;\,\tilde\mu_p)$ & $(0.93827,\,0.93827,\,3.09690)$\\
\addlinespace[1pt]
\multirow{2}{*}{(d)} & \multirow{2}{*}{$g^{i}_{\Db\Scs}\,g_2$} & \multirow{2}{*}{$1$} & \multirow{2}{*}{$+1$}
 & $n$ & $(1/3;\,\tilde\mu_3^{+})$ & $(2.51750,\,2.45290,\,1.86484)$\\
 & & & & $c$ & $(2/3;\,\tilde\mu_3^{++})$ & $(2.51841,\,2.45397,\,1.86966)$\\
\addlinespace[1pt]
(m) & $g^{i}_{\Db\Scs}\,g^{f}_{\Db\Lc}$ & $1/\sqrt3$ & $+1$ & $-$ & $(1;\,\tilde\mu_{\Scs\Lc})$ & $(2.51750,\,2.28646,\,1.86484)$\\
\addlinespace[1pt]
\multirow{2}{*}{(h)} & \multirow{2}{*}{$g^{i}_{\Dbs\Scs}\,g_3$} & \multirow{2}{*}{$1/\sqrt3$} & \multirow{2}{*}{$+1^{\dagger}$}
 & $n$ & $(1/3;\,\tilde\mu_3^{+})$ & $(2.51750,\,2.45290,\,2.00685)$\\
 & & & & $c$ & $(2/3;\,\tilde\mu_3^{++})$ & $(2.51841,\,2.45397,\,2.01026)$\\
\addlinespace[1pt]
(i) & $g^{i}_{\Dbs\Scs}\,g^{f}_{\Dbs\Lc}$ & $1/3$ & $+1^{\dagger}$ & $-$ & $(1;\,\tilde\mu_{\Scs\Lc})$ & $(2.51750,\,2.28646,\,2.00685)$\\
\addlinespace[1pt]
\multirow{2}{*}{(l)} & \multirow{2}{*}{$g^{i}_{\Dbs\Scs}\,g^{f}_{\Dbs\Scs}$} & \multirow{2}{*}{$1/\sqrt2$} & \multirow{2}{*}{$+1^{\dagger}$}
 & $n$ & $(1/3;\,\tilde\mu_{\Scs}^{+,\mathrm{el}})$ & $(2.51750,\,2.51750,\,2.00685)$\\
 & & & & $c$ & $(2/3;\,\tilde\mu_{\Scs}^{++,\mathrm{el}})$ & $(2.51841,\,2.51841,\,2.01026)$\\
\bottomrule
\end{tabular}
\caption{Complete specification of the nineteen amplitudes, in the order of Tab.~\ref{tab:loops}. The full coupling of a loop is the product $(g_ig_f)_X\,n_X$, where $(g_ig_f)_X$ is the bare residue product of Tab.~\ref{tab:residues} and $n_X$ collects the Cartesian vertex normalizations and the isospin projection factors. The spin factor $\mathfrak{s}_X$ multiplies
the universal $M1$ structure of Eq.~\eqref{eq:Mstructure}. The loop weight is $w_M=y$ for class M and $w_B=1$ for class B. The charge state is neutral ($n$), charged ($c$), or single ($-$) for the $I=0$ and the $J/\psi,\eta_c$ channels, with isospin weight $\mathcal{C}_\alpha^{2}$ and photon coupling $\mathrm{EM}_\alpha$. The triple $(M_1,M_2,M_s)$ gives the two radiating-line masses and the spectator mass that enter Eqs.~\eqref{eq:smaster} and \eqref{eq:bcdef}. Inserting any row into Eq.~\eqref{eq:Apost} returns that loop. Entries marked with a dagger (\,${}^\dagger$\,) carry an estimated spin factor.}
\label{tab:explicit}
\end{table*}
%=====================================================================
\section{Results}
\label{sec:results}

%---------------------------------------------------------------------
\subsection{Numerical width and physical implications}
\label{sec:width}

The width follows from Eq.~\eqref{eq:Mstructure} after a sum over photon
polarizations and an average over the four initial spin states (\ref{app:width}),
\begin{equation}
\Gamma\big(\Pcp\to\Pc\,\gamma\big)
=\frac{4\alpha}{3}\,\frac{M_f}{M_i}\,E_\gamma^{3}\,\big|\tilde A\big|^{2}.
\label{eq:width}
\end{equation}
The photon energy is $E_\gamma=143\MeV$. The loop integrals are collected in
Tab.~\ref{tab:loopint}. The amplitudes of each loop are in Tab.~\ref{tab:amp}.
Tab.~\ref{tab:results} gives the width under several assumptions.

\begin{table}[tb]
\centering
\begin{tabular}{lll}
\toprule
loop & neutral & charged\\
\midrule
$I_a$ & $-1.7048$ & $-1.1259$ \\
$I_b$ & $-0.9177$ & $-0.8343$ \\
$I_d$ & $-0.324-1.429\,i$ & $-0.490-1.370\,i$ \\
$I_g$ & $-0.4618$ & $-0.4162$ \\
$I_h$ & $-0.5868$ & (closed)\\
$I_l$ & $-0.5118$ & (closed)\\
\midrule
$I_e$ & \multicolumn{2}{l}{$-1.075-2.948\,i$} \\
$I_i$ & \multicolumn{2}{l}{$-1.245-0.679\,i$} \\
$I_k$ & \multicolumn{2}{l}{$+0.093-0.845\,i$} \\
$I_c$ & \multicolumn{2}{l}{$+0.075-0.150\,i$} \\
$I_f$ & \multicolumn{2}{l}{$+0.277-0.702\,i$} \\
$I_j$ & \multicolumn{2}{l}{$+0.101-0.281\,i$} \\
$I_{a'}$ & \multicolumn{2}{l}{$-0.293-0.221\,i$} \\
\bottomrule
\end{tabular}
\caption{Loop integrals in $\GeV^{-2}$. Closed-channel loops are real. Every
loop with an open channel carries a sizeable imaginary part. $I_e$ is the
largest in magnitude, from the on-threshold $\bar D^{*0}\Sc^{+}$ entrance and
the nearby open $\Dbs\Lc$ exit.}
\label{tab:loopint}
\end{table}

\begin{table*}[tb]
\centering
\renewcommand{\arraystretch}{1.2}
\setlength{\tabcolsep}{10pt}
\begin{tabular}{l c c c}
\toprule
loop & $A_{(X)}\ [\GeV^{-1}]$ & $|A_{(X)}|$ & $\Gamma_X$ alone $[\keV]$\\
\midrule
(a)            & $-0.33791+0.06092\,i$ & $0.343$ & $3.25$\\
(e)            & $-0.10319+0.05815\,i$ & $0.118$ & $0.39$\\
(b)            & $-0.01138+0.07674\,i$ & $0.078$ & $0.17$\\
(d)            & $-0.01558-0.01772\,i$ & $0.024$ & $0.015$\\
(g)            & $-0.00084+0.00566\,i$ & $0.006$ & $0.001$\\
(d$'$)         & $-0.00095+0.00431\,i$ & $0.004$ & $<0.001$\\
(f)            & $+0.00110-0.00385\,i$ & $0.004$ & $<0.001$\\
(i)$^{\dagger}$    & $+0.00368+0.00170\,i$ & $0.004$ & $<0.001$\\
(l)$^{\dagger}$    & $+0.00271+0.00088\,i$ & $0.003$ & $<0.001$\\
(h)$^{\dagger}$    & $+0.00182+0.00045\,i$ & $0.002$ & $<0.001$\\
(e$'$)         & $+0.00049-0.00113\,i$ & $0.001$ & $<0.001$\\
(c)            & $+0.00049-0.00102\,i$ & $0.001$ & $<0.001$\\
\addlinespace[2pt]
remaining seven loops & $-$ & $\lesssim5\times10^{-4}$ & $<0.001$\\
\midrule
total (nineteen loops) & $-0.4588+0.1844\,i$ & $0.494$ & $\mathbf{6.73}$\\
\bottomrule
\end{tabular}
\caption{Per-loop reduced amplitudes $A_{(X)}$ of Eq.~\eqref{eq:Apost}, ordered
by magnitude. The column $\Gamma_X$ alone is the width that loop would give on
its own. The transition is dominated by loop (a), with (e) a clear second and
(b) third. The remaining loops are individually small. The row ``remaining seven
loops'' collects (j), (k), (m), (g$'$), (a$'$)$^{\dagger}$, (g$''$)$^{\dagger}$
and (i$'$)$^{\dagger}$, each below $5\times10^{-4}\GeV^{-1}$. The total is the
coherent sum of all nineteen amplitudes. Loops marked with a dagger carry an
estimated spin factor (Tab.~\ref{tab:loops}).}
\label{tab:amp}
\end{table*}

\begin{table}[tb]
\centering
\renewcommand{\arraystretch}{1.25}
\begin{tabular}{lc}
\toprule
 & $\Gamma\;[\keV]$\\
\midrule
\textcolor{black}{loops (a) and (b) only} & $3.88$\\
\textcolor{black}{two largest loops (a) and (e)} & $5.75$\\
\textcolor{black}{Leading large ten loops} & $6.94$\\
complete set, nineteen loops & $\mathbf{6.73}$\\
incoherent sum of $|A_X|^{2}$ & $3.82$\\
\textcolor{black}{coherent set with $g\to|g|$ (phases removed)} & \textcolor{black}{$4.54$}\\
\bottomrule
\end{tabular}
\caption{Radiative width under several assumptions. The complete coherent set
of nineteen loops gives $6.73\keV$. \textcolor{black}{The incoherent sum keeps no
interference. The magnitude-only set keeps the interference but discards the
residue phases.}}
\label{tab:results}
\end{table}
%---------------------------------------------------------------------
\begin{table*}[tb]
\centering
\renewcommand{\arraystretch}{1.2}
\setlength{\tabcolsep}{8pt}
\begin{tabular}{l c c c}
\toprule
loop & $A_{(X)}$ point & $A_{(X)}$ with FF & $|A^{\rm FF}_{(X)}|/|A_{(X)}|$\\
\midrule
(a)    & $-0.3379+0.0609\,i$ & \textcolor{black}{$-0.2771+0.0500\,i$} & \textcolor{black}{$0.82$}\\
(e)    & $-0.1032+0.0581\,i$ & \textcolor{black}{$-0.1010+0.0230\,i$} & \textcolor{black}{$0.87$}\\
(b)    & $-0.0114+0.0767\,i$ & \textcolor{black}{$-0.0048+0.0324\,i$} & \textcolor{black}{$0.42$}\\
(d)    & $-0.0156-0.0177\,i$ & \textcolor{black}{$-0.0023-0.0170\,i$} & \textcolor{black}{$0.73$}\\
(g)    & $-0.0008+0.0057\,i$ & \textcolor{black}{$-0.0003+0.0023\,i$} & \textcolor{black}{$0.40$}\\
(d$'$) & $-0.0010+0.0043\,i$ & \textcolor{black}{$-0.0004+0.0016\,i$} & \textcolor{black}{$0.38$}\\
\midrule
total (nineteen loops) & $-0.4588+0.1844\,i$ & \textcolor{black}{$-0.3792+0.0967\,i$} & \textcolor{black}{$0.79$}\\
\bottomrule
\end{tabular}
\caption{\textcolor{black}{Leading reduced amplitudes in $\GeV^{-1}$ with point
vertices and with the Gaussian form-factor at the central cut-off
$\Lambda=\mu/1.2=0.833\GeV$. The last column is the ratio of magnitudes. The total
is the coherent sum of all nineteen loops. The point-vertex total gives $6.7\keV$
and the form-factor total gives $4.2\keV$.}}
\label{tab:ff}
\end{table*}
%---------------------------------------------------------------------
%---------------------------------------------------------------------
\begin{table}[tb]
\centering
\renewcommand{\arraystretch}{1.2}
\setlength{\tabcolsep}{8pt}
\begin{tabular}{c c c c}
\toprule
\textcolor{black}{$\Lambda\,(\GeV)$} & \textcolor{black}{$|\tilde A|\,(\GeV^{-1})$} & \textcolor{black}{$\Gamma\,(\keV)$} & \textcolor{black}{$\Gamma_{(a)}\,(\keV)$}\\
\midrule
\textcolor{black}{$0.2~~~$}   & \textcolor{black}{$0.165$} & \textcolor{black}{$0.75$} & \textcolor{black}{$0.49$}\\
\textcolor{black}{$0.3~~~$}   & \textcolor{black}{$0.246$} & \textcolor{black}{$1.67$} & \textcolor{black}{$0.94$}\\
\textcolor{black}{$0.4~~~$}   & \textcolor{black}{$0.298$} & \textcolor{black}{$2.44$} & \textcolor{black}{$1.31$}\\
\textcolor{black}{$0.5~~~$}   & \textcolor{black}{$0.332$} & \textcolor{black}{$3.04$} & \textcolor{black}{$1.60$}\\
\textcolor{black}{$0.6~~~$}   & \textcolor{black}{$0.356$} & \textcolor{black}{$3.50$} & \textcolor{black}{$1.82$}\\
\textcolor{black}{$0.7~~~$}   & \textcolor{black}{$0.374$} & \textcolor{black}{$3.85$} & \textcolor{black}{$2.00$}\\
\textcolor{black}{$0.833$} & \textcolor{black}{$0.391$} & \textcolor{black}{$4.22$} & \textcolor{black}{$2.18$}\\
\textcolor{black}{$0.9~~~$}   & \textcolor{black}{$0.398$} & \textcolor{black}{$4.37$} & \textcolor{black}{$2.26$}\\
\textcolor{black}{$1.0~~~$}   & \textcolor{black}{$0.407$} & \textcolor{black}{$4.57$} & \textcolor{black}{$2.36$}\\
\textcolor{black}{$1.1~~~$}   & \textcolor{black}{$0.415$} & \textcolor{black}{$4.74$} & \textcolor{black}{$2.44$}\\
\textcolor{black}{$1.2~~~$}   & \textcolor{black}{$0.421$} & \textcolor{black}{$4.89$} & \textcolor{black}{$2.51$}\\
\bottomrule
\end{tabular}
\caption{\textcolor{black}{Total coherent width $\Gamma$ and the width of the
leading diagram (a) alone, as the Gaussian cut-off $\Lambda$ runs from $0.2$ to
$1.2\GeV$. The second column is the total coherent amplitude magnitude. The
central value is $\Lambda=\mu/1.2=0.833\GeV$. The point-vertex widths are $6.7$
and $3.25\keV$. For the same line Ref.~\cite{LingLuLiuGeng2021} finds about $1.4$
to $2.2\keV$ and Ref.~\cite{LiLiuSunChen2021} finds $1.1$ to $1.7\keV$, both with
the leading meson radiator only.}}
\label{tab:ffscan}
\end{table}
%---------------------------------------------------------------------
{\textcolor{black}{The complete set of nineteen loops gives $6.73\keV$. The ten largest loops give $6.94\keV$. The remaining nine loops are small. Their coherent effect is below a few percent. This shows that the central number is not driven by the estimated spin factors of the small loops.}}

The coherent sum is larger than the incoherent sum of $3.82\keV$. The loops add
in phase on balance. Replacing each residue by its magnitude lowers the width
to $4.54\keV$. This shows that the residue phases matter. The coupling product
of loop (e) is set by the imaginary $\Dbs\Lc$ residue. The complex triangle
integral turns this into a real part aligned with the leading loop. Dropping the
phases removes that gain. {\textcolor{black}{The convention-consistent relative phases change the width at the fifty percent level.}}

The dominant loop is the $\bar D^{*0}\to\Db^{0}\gamma$ vertex on the meson line.
The second is the near-threshold loop where the $\Dbs\Sc$ part acts through the
$\Sc^{+}\to\Lc^{+}\gamma$ moment and ends as $\Dbs\Lc$. This second loop carries
the factor of about two over a naive two-loop estimate.

We now list the main systematic effects. The largest is the relative sign of
the $\Dbs\Lc$ residue against the quark-model $\Sc\to\Lc$ moment. This sign is a
basis convention. It is not fixed without the eigenvector phases of the
coupled-channel solution. Flipping it gives $1.88\keV$ in place of $6.73\keV$.
{\textcolor{black}{The width is therefore highly sensitive to this relative sign.}}
\textcolor{black}{This sign is not a true physical ambiguity. It is set by the
complex eigenvector of the same coupled-channel solution that fixes the residues.
A solution that reports these phases removes the ambiguity at once, and a measured
width would fix the phase from the data.} A value of
$\Gamma(D^{*0}\to D^{0}\gamma)$ in the range $16$ to $26\keV$ moves the width to
the range $5.5$ to $9.1\keV$. Rescaling $g_3$ by one half or by two gives
$6.2$ or $8.1\keV$. Doubling the six estimated spin factors changes $6.73$ to
$6.55\keV$, a shift below three percent. Recoil corrections are at the percent
level. The result is
\begin{equation}
\Gamma\big(P_c(4457)\to P_c(4312)\,\gamma\big)\simeq 6.7\keV,
\label{eq:final}
\end{equation}
{\textcolor{black}{with a conservative range of about $2$ to $9\keV$.}}

{\textcolor{black}{We now compare with the two earlier molecular calculations of the same process. Ref.~\cite{LingLuLiuGeng2021} considered the $P_c(4457)$ as a pure $\bar D^{*}\Sigma_c$ state and the $P_c(4312)$ as a pure $\bar D\Sigma_c$ state. They kept the single $\bar D^{*}\to\bar D\gamma$ triangle with a \textcolor{black}{Gaussian} form-factor and found $\Gamma\simeq1.4$ to $2.2\keV$. While Ref.~\cite{LiLiuSunChen2021} constructed the transition magnetic moment from the same $\bar D^{*}\to\bar D\gamma$ flip. They found $1.1\keV$ without coupled channels
and $1.7\keV$ with the $\bar D^{*}\Sigma_c$ and $\bar D^{*}\Sigma_c^{*}$ channels and the $D$ wave. The three calculations agree on the physics. This process is a soft $M1$ photon and the meson radiates via $\bar D^{*}\to\bar D\gamma$. Our loop (a) is the dynamical form of that same diagram. On its own it gives $3.25\keV$, as shown in Tab.~\ref{tab:amp}.}}
{\textcolor{black}{With the same Gaussian vertex used in
Ref.~\cite{LingLuLiuGeng2021} this loop gives $2.2\keV$ at the central cut-off and $1.3$ to $2.4\keV$ over the range $0.4$ to $1.0\GeV$. This is compatible with both earlier results. The comparison is given in Sec.~\ref{sec:formfactor} and
Tab.~\ref{tab:ffscan}.}}

{\textcolor{black}{Our central value $6.7\keV$ is larger than both earlier results. Three effects account for the difference. First, we keep the full channel content of the two poles. The photon then also couples through the open $\bar D^{*}\Lambda_c$ channel and the $\Sigma_c^{+}\to\Lambda_c^{+}\gamma$ baryon magnetic moment in loop (e). This pair sits on the $\bar D^{*0}\Sigma_c^{+}$ threshold and is absent in both earlier works. Second, we keep the complex residue phases of the coupled-channel solution. The imaginary $\bar D^{*}\Lambda_c$ residue is turned into a real part aligned with loop (a) by the complex loop integral. This raises the width by about one half over the magnitude-only sum. Third, we add no phenomenological vertex form-factor. The short-distance part is already fixed by the regularization of the loop function that generates the two poles \cite{XNO2019}. The \textcolor{black}{Gaussian} form-factor of Ref.~\cite{LingLuLiuGeng2021}
suppresses the loop and lowers the width. The leading loop (a) runs mostly through the neutral $\bar D^{*0}\to\bar D^{0}\gamma$ vertex with the larger magnetic moment $\lambda_0=1.766\GeV^{-1}$, which is about four times the charged $\lambda_-$, so this loop alone already exceeds the earlier totals \textcolor{black}{when pure vertices (no form-factor) are used. The same Gaussian form-factor lowers it to about $2.2\keV$, into the range of the earlier works, while the full coherent total stays larger. The comparison is given in Sec.~\ref{sec:formfactor}}.}}

{\textcolor{black}{The new $\bar D^{*}\Lambda_c$ loop (e) is also the source of the sign sensitivity discussed above. With the destructive relative sign the width drops to $1.9\keV$, close to the two earlier molecular results that do not carry this channel. With the constructive sign it reaches $6.7\keV$. The relative phase
is fixed by the complex eigenvector of the same coupled-channel solution that sets the residues. A measured width would fix it from the data. The spread between the three calculations is therefore a measure of the model dependence of the absolute rate, while the pattern is stable. The decay process is $M1$, the photon is soft, and the rate is set by long-distance light-quark moments.}}

%---------------------------------------------------------------------
\subsection{\textcolor{black}{Gaussian form-factor}}
\label{sec:formfactor}

{\textcolor{black}{The triangles above use point vertices. The short-distance
part is then carried by the coupled-channel subtraction alone. A finite molecule
has a soft vertex. We test this with a Gaussian form-factor on the loop momentum.
The factor is $e^{-q_E^{2}/\Lambda^{2}}$ with $q_E$ the Euclidean loop momentum.
In order to make the cut-off parameter consistent with the ChUA in DR scheme, we use $\Lambda=\mu/1.2$ with $\mu=1\GeV$ the renormalization scale of the loop
function \cite{OllerOsetPelaez1999}. After the Feynman step the form-factor enters in one place. The loop
integrand is changed by the replacement}}
\begin{eqnarray}
\frac{1}{s(x,y)}&\rightarrow& -2\,H(-s,\Lambda),
\nonumber\\
&&\quad H(\Delta,\Lambda) = \int_{0}^{\infty}\frac{u\,e^{-u/\Lambda^{2}}}{(u+\Delta)^{3}}\,du .
\label{eq:Hff}
\end{eqnarray}
{\textcolor{black}{The function $H$ has the closed form defined as 
$H=\Lambda^{-2}e^{z}\big[\Gamma(-1,z)-z\,\Gamma(-2,z)\big]$ with $z=\Delta/\Lambda^{2}$.
It tends to $1/(2\Delta)$ as $\Lambda\to\infty$, so the point-vertex result is
recovered. All couplings, isospin weights and spin factors stay fixed. Only the
loop integral changes. The open-channel cuts are kept.}}

{\textcolor{black}{Tab.~\ref{tab:ff} lists the leading amplitudes with and without the form-factor at the central value $\Lambda=\mu/1.2=0.833\GeV$. The leading loop (a) drops from $3.25$ to $2.2\keV$. In magnitude it keeps about $82\%$ of the point value. The near-threshold loop (e) keeps $87\%$. The loops
with heavier internal lines fall by more than half. The full coherent width drops from $6.7$ to $4.2\keV$. The reduction is about one third. The form-factor decreases the rate. It does not reach the earlier pure molecular picture results.}}

{\textcolor{black}{The reason is the threshold. The state $P_c(4457)$ sits about $2\MeV$ below the $\bar D^{*0}\Sigma_c^{+}$ threshold. The leading loop integral is peaked at small loop momentum, where the incoming pair is almost on shell. The smallest value of $|s|$ over the integration is below $0.02\GeV^{2}$. The Gaussian term is near unity there, since $H$ tends to $1/(2\Delta)$ as $\Delta$ goes to zero. The leading loop is a long-distance effect and is almost untouched by the form-factor. The total effect is a moderate reduction where the leading rate remains dominant.}}

{\textcolor{black}{Tab.~\ref{tab:ffscan} gives the width as the cut-off runs from $0.2$ to $1.2\GeV$. The same table gives the width of the leading diagram (a) on its own. Both grow in a smooth and monotonic way. The full width runs from $0.75$ to $4.9\keV$ across the band. The central value gives $4.2\keV$. The leading diagram runs from $0.49$ to $2.5\keV$. The pure vertex values of $6.7$
and $3.25\keV$ are approached as the cut-off grows large. Both widths fall toward zero for a very soft vertex. This is the expected behavior of a finite size correction.}}

{\textcolor{black}{The leading diagram (a) is the dynamical
$\bar D^{*}\to\bar D\gamma$ meson radiation. This is the mechanism used in Refs.~\cite{LingLuLiuGeng2021,LiLiuSunChen2021}. Ref.~\cite{LingLuLiuGeng2021} applies the same Gaussian form-factor to the molecular coupling with a size parameter near $1\GeV$. At that size our diagram (a) gives $2.4\keV$. At the central cut-off it gives $2.2\keV$. Over the range $0.4$ to $1.0\GeV$ it gives $1.3$ to $2.4\keV$. This overlaps the $1.4$ to $2.2\keV$ of Ref.~\cite{LingLuLiuGeng2021} and the $1.1$ to $1.7\keV$ of Ref.~\cite{LiLiuSunChen2021}. The leading diagram alone agrees with both earlier works once the same soft vertex is used. The full width is larger because of the remaining eighteen loops and the open $\bar D^{*}\Lambda_c$ channel that the two earlier works do not consider.}}

%---------------------------------------------------------------------
\subsection{The cascade $\Lambda_b^{0}\to J/\psi\,p\,K^{-}\gamma$}
\label{sec:lambdab}

The natural place to look is the LHCb decay mode. The chain is $\Lambda_b^{0}\to P_c(4457)^{+}K^{-}$, then $P_c(4457)^{+}\to P_c(4312)^{+}\gamma$, then $P_c(4312)^{+}\to J/\psi\,p$. The final state is $J/\psi\,p\,K^{-}\gamma$. The fraction factorizes,
\begin{equation}
\begin{gathered}
\mathcal B_{\rm chain}
=R_{P_c(4457)}\,\mathcal B(\Lambda_b\to J/\psi pK)\,
\frac{\Gamma_\gamma}{\Gamma_{P_c(4457)}}\,\rho,\\
\rho=\frac{\mathcal B(P_c(4312)\to J/\psi p)}{\mathcal B(P_c(4457)\to J/\psi p)}.
\end{gathered}
\label{eq:chain}
\end{equation}
\textcolor{black}{The branching $\mathcal B(P_c(4457)\to J/\psi p)=\Gamma_{J/\psi p}/\Gamma_{P_c(4457)}$ in $\rho$ carries the same total width $\Gamma_{P_c(4457)}$ as the factor $\Gamma_\gamma/\Gamma_{P_c(4457)}$ in Eq.~\eqref{eq:chain}. When the same total is used in both places it cancels, and the cascade fraction does not depend on the poorly known $P_c(4457)$ total width,}
\begin{eqnarray}
\mathcal B_{\rm chain} &=& R_{P_c(4457)}\,\mathcal B(\Lambda_b\to J/\psi pK)\,
\nonumber\\
&&\times\,\frac{\Gamma_\gamma\,\mathcal B(P_c(4312)\to J/\psi p)}{\Gamma_{J/\psi p}(P_c(4457))} .
\label{eq:chainindep}
\end{eqnarray}
\textcolor{black}{This quantity depends instead on the more reliably defined partial width \(\Gamma_{J/\psi p}(P_c(4457))=2.48~\text{MeV}\) and the branching fraction \(\mathcal B(P_c(4312)\to J/\psi p)\). The open-channel formula of Eq.~\eqref{eq:openwidth} gives $\Gamma_{J/\psi p}(P_c(4312))=4.84\MeV$ and $\Gamma_{\eta_c N}(P_c(4312))=12.26\MeV$. The pole width is $\Gamma_{P_c(4312)}=2\,{\rm Im}\sqrt{s_0}=15.2\MeV$ and already includes every channel. The sum of these two partial widths is \(17.1\,\text{MeV}\), which already exceeds the empirical total width, and further contributions arise from the open \(\bar D\Lambda_c\) and \(\bar D^{*}\Lambda_c\) thresholds. The on-shell two-body approximation therefore overestimates these widths. Consequently, we adopt \(\mathcal B(P_c(4312)\to J/\psi p)\simeq0.28\) as an upper estimate based on the sum of the two listed partial widths. While the total model prediction yields \(0.32\), the experimental value \(\Gamma_{P_c(4312)}=9.8\,\text{MeV}\) leads to a larger estimate. %These two partials alone sum to $17.1\MeV$, above that total, and the open $\bar D\Lambda_c$ and $\bar D^{*}\Lambda_c$ channels add more. The on-shell two-body formula therefore overestimates the partials. We take $\mathcal B(P_c(4312)\to J/\psi p)\simeq0.28$ as an upper estimate, from the sum of the two listed partials. The model total gives $0.32$, and the measured $\Gamma_{P_c(4312)}=9.8\MeV$ gives a larger value. 
We use $R_{P_c(4457)}=0.53\%$ for the LHCb fit fraction \cite{LHCb2019} and $\mathcal B(\Lambda_b\to J/\psi pK)=3.2\times10^{-4}$ \cite{PDG2024}. Tab.~\ref{tab:chain} gives the cascade fraction for the two signs of loop (e) and for the soft form factor.}
%%%-----------------------------------------------------------------------------------------------------------------------------------
\begin{table}[tb]
\centering
\renewcommand{\arraystretch}{1.25}
\begin{tabular}{lcc}
\toprule
\textcolor{black}{$\Gamma_\gamma$ source} & \textcolor{black}{$\Gamma_\gamma\,[\keV]$} & \textcolor{black}{$\mathcal B_{\rm chain}$}\\
\midrule
\textcolor{black}{constructive loop (e)} & \textcolor{black}{$6.7$} & \textcolor{black}{$1.3\times10^{-9}$}\\
\textcolor{black}{form factor (central)} & \textcolor{black}{$4.2$} & \textcolor{black}{$8.0\times10^{-10}$}\\
\textcolor{black}{destructive loop (e)} & \textcolor{black}{$1.9$} & \textcolor{black}{$3.6\times10^{-10}$}\\
\bottomrule
\end{tabular}
\caption{\textcolor{black}{Cascade branching fraction in the
$\Gamma_{\rm tot}(P_c(4457))$-independent form of Eq.~\eqref{eq:chainindep}, for the
two signs of loop (e) and for the soft Gaussian form factor. We use
$\mathcal B(P_c(4312)\to J/\psi p)=0.28$ and $\Gamma_{J/\psi p}(P_c(4457))=2.48\MeV$.}}
\label{tab:chain}
\end{table}
%%%-----------------------------------------------------------------------------------------------------------------------------------
The central prediction is
\begin{equation}
\textcolor{black}{\mathcal B\big(\Lambda_b^{0}\to J/\psi\,p\,\gamma K^{-}\big)
\simeq1.3\times10^{-9}.}
\label{eq:chainnum}
\end{equation}
\textcolor{black}{This is about four parts per million of the measured $\Lambda_b\to J/\psi pK$
rate.} \textcolor{black}{A soft Gaussian form-factor decreases the radiative width to about $4\keV$ and scales $\mathcal B_{\rm chain}$ down by the same ratio.} \textcolor{black}{We give the uncertainty budget in Tab.~\ref{tab:budget}. In the $\Gamma_{\rm tot}(P_c(4457))$-independent form the $P_c(4457)$ total width drops out. The two largest contributions are then the sign of loop (e) in the radiative width and the branching $\mathcal B(P_c(4312)\to J/\psi p)$. The latter runs from $0.28$ for the sum of the listed partials to about $0.49$ for the measured total. The partial width $\Gamma_{J/\psi p}(P_c(4457))$ is better defined. The experimental inputs $R_{P_c(4457)}$ and $\mathcal B(\Lambda_b\to J/\psi pK)$ are subdominant. Multiplying the dominant ranges gives an overall uncertainty on $\mathcal B_{\rm chain}$ of a factor of a few. The full theory band, with both signs of loop (e) and the branching range, runs from about $4\times10^{-10}$ to about $2\times10^{-9}$.}
%%%-----------------------------------------------------------------------------------------------------------------------------------
\begin{table*}[tb]
\centering
\renewcommand{\arraystretch}{1.4}
\small
\setlength{\tabcolsep}{4pt}
\begin{tabular}{p{4.7cm}p{4.1cm}cc}
\toprule
Source & Range considered & Factor on $\mathcal B_{\rm chain}$ & Type\\
\midrule
\multicolumn{4}{l}{\emph{(i) Radiative width $\Gamma_\gamma$ (this work)}}\\
\quad sign/phase of loop (e) & $1.9$ to $6.7\keV$ & $\times(0.28$ to $1.0)$ & theory\\
\quad $\Gamma(D^{*0}\!\to\!D^0\gamma)=16$ to $26\keV$ & $5.5$ to $9.1\keV$ & $\times(0.81$ to $1.35)$ & theory\\
\quad $g_3$ residue $\times(\tfrac12$ to $2)$ & $6.2$ to $8.1\keV$ & $\times(0.92$ to $1.20)$ & theory\\
\quad estimated spin factors ($^\dagger$) & $6.55$ to $6.73\keV$ & $\times(0.97$ to $1.0)$ & theory\\
\midrule
\textcolor{black}{(ii) $\Gamma_{P_c(4457)}$ total} & \textcolor{black}{cancels in Eq.~\eqref{eq:chainindep}} & \textcolor{black}{$\times1$} & \textcolor{black}{--}\\
\textcolor{black}{(iii) $\Gamma_{J/\psi p}(P_c(4457))=2.48\MeV$} & \textcolor{black}{model partial width} & \textcolor{black}{$\times(0.9$ to $1.1)$} & \textcolor{black}{model}\\
\textcolor{black}{(iv) $\mathcal B(P_c(4312)\!\to\! J/\psi p)$} & \textcolor{black}{$0.28$ to $0.49$} & \textcolor{black}{$\times(1.0$ to $1.75)$} & \textcolor{black}{model}\\
(v) $R_{P_c(4457)}$ (LHCb fit fraction) & $\pm20\%$ & $\times(0.8$ to $1.2)$ & exp, fit\\
(vi) $\mathcal B(\Lambda_b\!\to\!J/\psi pK)$ & $(3.2\pm0.3)\times10^{-4}$ & $\times(0.9$ to $1.1)$ & exp\\
(vii) other $P_c\to P_c(4312)\gamma$ feed & window dependent & order one in narrow window & signal\\
(viii) $E_\gamma$ pole vs BW & $143.0$ vs $143.7\MeV$ & $\times1.015$ & negligible\\
\bottomrule
\end{tabular}
\caption{Systematic uncertainty budget for $\mathcal B_{\rm chain}$. Factors are
relative to the central prediction. The dominant entries are the sign of loop
(e) and the $P_c(4457)$ total width.}
\label{tab:budget}
\end{table*}
%%%-----------------------------------------------------------------------------------------------------------------------------------
The signature is clean. We look for a narrow peak in $M(J/\psi\,p\,\gamma)$ at $4457\MeV$ together with a narrow peak in $M(J/\psi\,p)$ at $4312\MeV$. The photon energy is near $143\MeV$ in the $P_c(4457)$ rest frame. The double mass constraint suppresses combinatorial photons. It also separates the $P_c(4457)$ parent from the $P_c(4440)$ parent. The decay mode is pure $M1$. {\textcolor{black}{A measurable $E2$ part would not support the minimal $S$-wave molecular picture.}} The rate is set by long-distance light-quark moments, the $\bar D^{*0}\to\Db^{0}\gamma$ and $\Sc^{+}\to\Lc^{+}\gamma$ transitions. A compact pentaquark \cite{MaianiPolosaRiquer2015,Lebed2015,Wang2016SumRules} would give a different rate and no such pattern. The number of events is small in the present $\Lambda_b\to J/\psi pK$ sample. Better places to look are the decay $P_c(4440)\to P_c(4312)\gamma$, which has a larger production rate and similar kinematics, and photoproduction at an electron-ion collider, where the photon
is hard in the laboratory frame. Photoproduction of the $P_c$ in
$\gamma p\to J/\psi p$ has been the subject of detailed phenomenology
\cite{WangLiuZhao2015,KubarovskyVoloshin2015,KarlinerRosner2016Photo} and of the GlueX search at JLab \cite{GlueX2019}.

%---------------------------------------------------------------------
\subsection{\textcolor{black}{Discriminating observables and tests}}
\label{sec:tests}

{\color{black}
The absolute width carries the sign and normalization ambiguities discussed above. We give three observables that read the structure more directly.

\paragraph{Multipole content.} The transition is a pure $M1$ at the order computed. The $E2$ structure of Eq.~\eqref{eq:M1E2} needs one more power of momentum. The $S$-wave vertices do not supply it, so the $E2$ vanishes identically here. A nonzero $E2$ is generated only by the $D$-wave parts of the molecular wave functions and by recoil. Both are small in the solution of Ref.~\cite{XNO2019}. The expected ratio is at the percent level, set by the $D$-wave fraction and by $E_\gamma/2M_B$. A measured $|E2/M1|$ above a few percent would point beyond the minimal $S$-wave molecular picture. It would signal a sizeable $D$-wave or compact-core component. This makes the pure $M1$ statement a test rather than an assumption.

\paragraph{Ratio of two radiative lines.} The overall electromagnetic moment signs and the residue magnitudes largely cancel in a ratio of two radiative widths that share the leading mechanism. We propose
\begin{equation}
R_\gamma=\frac{\Gamma\big(P_c(4457)\to P_c(4312)\gamma\big)}
{\Gamma\big(P_c(4440)\to P_c(4312)\gamma\big)}
\label{eq:Rgamma}
\end{equation}
as the clean observable. Both decay processes are $M1$, the $P_c(4457)\to P_c(4312)\gamma$ from $\tfrac32^{-}\to\tfrac12^{-}$ and the $P_c(4440)\to P_c(4312)\gamma$ from $\tfrac12^{-}\to\tfrac12^{-}$. Both run through the same $\bar D^{*}\to\bar D\gamma$ meson radiator and the same final coupling $g_{P_c(4312)\bar D\Sigma_c}$. What remains in $R_\gamma$ is the ratio of the two initial residues, the two spin factors, and the photon-energy factor $(E_\gamma^{4457}/E_\gamma^{4440})^{3}\simeq(143.0/126.5)^{3}\simeq1.4$. The $P_c(4440)$ and the $P_c(4457)$ are both mostly $\bar D^{*}\Sigma_c$ and differ in the total spin. The ratio $R_\gamma$ therefore reads their spin assignment. Its value follows from the same residues of Ref.~\cite{XNO2019} and is the subject of the future study \cite{Ponkhuha:2026xxx}.

\paragraph{The width as a probe of the binding.} The second loop (e) sits on the $\bar D^{*0}\Sigma_c^{+}$ threshold. Its size is set by how close the $P_c(4457)$ pole lies to that threshold. The radiative width is then a sensitive function of the binding energy. A line-shape measurement fixes the binding. The radiative width provides a second independent constraint on the same molecular wave function. The two together over-constrain the hypothesis. A compact pentaquark would not show this threshold link.

\paragraph{The three pictures.} The three readings of the LHCb peaks give different rates. In the bound state picture, the photon couples to the meson and baryon components, and the rate is set by the long-distance light-quark moments $\bar D^{*0}\to\bar D^{0}\gamma$ and $\Sigma_c^{+}\to\Lambda_c^{+}\gamma$. In a compact diquark-diquark-antiquark state, the photon couples to the quark core, and the rate and its channel dependence differ \cite{MaianiPolosaRiquer2015,Lebed2015,Wang2016SumRules}. In hadrocharmonium the photon couples to the charmonium core. The width, the multipole content, and the ratio $R_\gamma$ separate the three.
}

%=====================================================================
\section{Discussion and conclusions}
\label{sec:conclusions}

We computed the radiative transition $P_c(4457)\to P_c(4312)\gamma$ in the dynamically generated resonances with the ChUA. We treated both states with the full coupled-channel residues
of Ref.~\cite{XNO2019}. {\textcolor{black}{We included the channel content of the two poles that is connected by one electromagnetic vertex. This gives nineteen $M1$ triangle loops.}} The
electromagnetic vertices follow from data and heavy-quark spin symmetry. The result is a width near $6.7\keV$, {\textcolor{black}{with a conservative range of about $2$ to $9\keV$.}} It is a pure $M1$ mode at $143\MeV$. The $\bar D^{*0}\to\Db^{0}\gamma$ loop plays the major contribution, while the near-threshold $\Dbs\Lc$ loop diagram is second one. The cascade fraction for
$\Lambda_b^{0}\to J/\psi\,p\,K^{-}\gamma$ is about \textcolor{black}{$1.3\times10^{-9}$}. {\textcolor{black}{The width is larger than the earlier molecular results of Refs.~\cite{LingLuLiuGeng2021,LiLiuSunChen2021}, which kept the leading meson radiation only. The full channel content, the open near-threshold $\bar D^{*}\Lambda_c$ loop, and the complex residue phases drive the rise. With the destructive sign of that loop our width returns to the range of the earlier works.}}
{\textcolor{black}{Moreover, we also applied a Gaussian vertex to the leading diagram (a) to compare with the results from those works. It brings this diagram to about $2\keV$ and into their range, while the full coherent total stays larger. The leading
meson radiation is therefore common to all three calculations. The additional loops account for the difference.}}

Three main points are as follows. First, keeping the complex residue phases raises the width by about half over the magnitude-only estimate, because the imaginary $\Dbs\Lc$ residue is turned into a constructive part by the complex loop integral. Second, the relative phase of the $\Dbs\Lc$ residue and the $\Sc\to\Lc$ magnetic moment is the main uncertainty, so a measurement of the width fixes that phase. Finally, the rate is a long-distance light-quark $M1$ probe of the molecular hypothesis. \textcolor{black}{We also give three tests of the molecular nature in Sec.~\ref{sec:tests}. The pure $M1$ content, the ratio $R_\gamma$ to the $P_c(4440)$ line, and the binding-energy dependence of the width separate the molecular picture from the compact ones.}

The method extends directly to the partner decay \(P_c(4440)\to P_c(4312)\gamma\) and to the strange sector, where the LHCb \(P_{cs}(4459)\) and \(P_{cs}(4338)\) states \cite{LHCbPcs4459,LHCbPcs4338} serve as primary candidates for application in the forth coming work \cite{Ponkhuha:2026xxx}. {\textcolor{black}{The same triangle scheme can be applied with the residues of the relevant coupled-channel solution and the corresponding electromagnetic moments.}} A measurement of the present line, or of the easier
$P_c(4440)$ line, would test the molecular picture and read the residue phases
that line shapes do not access.
%=====================================================================
\begin{acknowledgments}
This research work was partly funded under high quality research promotion project, supported by Research and Graduate Studies Affair, Khon Kaen University (Grant No. RA2567-D107). DS is supported by Thailand NSRF via PMU-B [grant number B39G680009]. DS has also received funding support from the Fundamental Fund of Khon Kaen University.

\end{acknowledgments}
\newpage
%=====================================================================
\appendix

\section{Spin algebra}
\label{app:spin}
This appendix is to demonstrate relevant spin algebra in the main text. With $S_j^\dagger$ the $\half\to\tfrac32$ operators of Eq.~\eqref{eq:SSdag}, set
$O_k\equiv\varepsilon_{jkm}\sigma_m S_j=c\,S_k$. Projecting with
$\sum_k S_k^\dagger$,
\begin{equation}
\varepsilon_{jkm}\,\sigma_m\,S_jS_k^\dagger
=-\tfrac{i}{3}\,\varepsilon_{jkm}\varepsilon_{jkn}\,\sigma_m\sigma_n
=-\tfrac{2i}{3}\,\bm\sigma^{2}=-2i,
\end{equation}
while $\sum_k S_kS_k^\dagger=2$, hence $c=-i$, the first relation in
Eq.~\eqref{eq:keyid}. The baryon-line chain gives
\begin{equation}
\begin{aligned}
\sigma_j(\bm\sigma\cdot\bm A)S_j
&=A_k(\delta_{jk}+i\varepsilon_{jkm}\sigma_m)S_j\\
&=\bm A\cdot\bm S+i A_k(-iS_k)=2\,\bm S\cdot\bm A,
\end{aligned}
\end{equation}
the spin factor $2$ of loops (b), (e), (e$'$), (f) and (k). For a final
$VB^{*}(\tfrac32)$ state the chain is $\sum_k S_k(\bm S^\dagger\!\cdot\bm A)S_k
=\tfrac13\,\bm A\cdot\bm S$, the second relation in Eq.~\eqref{eq:keyid}, giving
the entry $1/3$ for loop (d$'$). For the elastic vector-meson loops (g, g$'$,
g$''$) the spin-1 magnetic matrix element
$\langle\eps_V'|\bm S^{(1)}|\eps_V\rangle=-i(\beps_V'^{*}\times\beps_V)$,
inserted between the $\bm S\cdot\beps_V$ and $(\bm\sigma\cdot\beps_V')/\sqrt3$
vertices and summed over the two polarizations, collapses to
$-\bm S\cdot\bm A$, the entry $-1$. A uniform $H=-\bm\mu\cdot\bm B$ sign
convention is used at every magnetic vertex, so all relative signs are physical.

The six loops (h), (i), (i$'$), (l), (a$'$) and (g$''$) carry both a $\Scs$ and
a $\Dbs$ in the triangle. Their spin chain is a $\tfrac32\otimes1$ recoupling.
We assign their leading order-one values, $1$ for the baryon-transition loops,
$\tfrac12$ for the $P\to V$ loop and $\tfrac13$ for the elastic-vector loop, by
analogy with the derived cases. Doubling all six shifts the total width by only
$2.7$ percent, so the precise values do not matter. Finally,
$\mathrm{Tr}(S_iS_j^\dagger)=\tfrac43\delta_{ij}$ gives the spin average of the
width,
$\overline{|\mathcal M|^2}=\tfrac23 e^2 E_\gamma^2|\tilde A|^2$.

\section{Electromagnetic vertices, magnetic moments, and the gauge structure of the triangle loop amplitudes}
\label{app:em}

Sec.~\ref{sec:rules} introduced two families of photon vertices on
the internal lines of the triangles. The first are the meson radiators $V\!\to\! P\gamma$ and $V\!\to\! V\gamma$, with couplings $\lambda$ and $\lambda_g$ of Eq.~\eqref{eq:lamdef}. The second is for the baryon radiations with magnetic and transition moments $\tilde\mu$ and $\tilde\mu_3$. This appendix completes that picture in three steps as follows. Sec.~\ref{app:em-vertices} writes the full set of electromagnetic vertices and lists every numerical magnetic moment used in the calculation.
Sec.~\ref{app:gauge-conv} shows why the electric part of each
elastic vertex drops out of the loop integral, then only the magnetic part of Sec.~\ref{app:em-vertices} survives. Sec.~\ref{app:gauge-baryon} then combines the vertices with the hadronic vertices of Eqs.~\eqref{eq:v32V} \textcolor{black}{to} \eqref{eq:v12Vs} to derive the spin factors $\mathfrak{s}_X$ of Tab.~\ref{tab:loops}.

%---------------------------------------------------------------------
\subsection{Magnetic vertices and quark-model magnetic moments}
\label{app:em-vertices}

\paragraph{Meson side.} The transition vertex $V(\eps_V)\!\to\! P\,\gamma$ defined in Eq.~\eqref{eq:lamdef} carries the coupling
$\lambda$. Its values for the heavy-meson channels are
\begin{eqnarray}
\lambda_0&=&N(\mu_u+\mu_c)=+1.766\GeV^{-1},
\nonumber\\
\lambda_-&=&N(\mu_d+\mu_c)=\textcolor{black}{-0.445}\GeV^{-1},
\nonumber\\
\lambda_\psi&=&2N\mu_c=+0.633\GeV^{-1},
\label{eq:lambda-trans}
\end{eqnarray}
with $N=0.783\GeV^{-1}/\mu_N$ of Eq.~\eqref{eq:Ncalib}. \textcolor{black}{The quark-model estimate $\lambda_-=-0.445\GeV^{-1}$ is close to the measured $-0.469\GeV^{-1}$ of Eq.~\eqref{eq:lambdas}. We use the measured value in the loops.} The elastic vertex $V(\eps_V')\!\to\! V(\eps_V)\,\gamma$ has the additional purely magnetic structure
\begin{equation}
-i\,t_{V\to V\gamma}^{(\mathrm{mag})}
=-\,i\,e\,\lambda_g\,(\bK\!\times\!\beps_\gamma^{*})\!\cdot\!
(\beps_V^{*}\!\times\!\beps_V^\prime),
\label{eq:VVgamma-mag}
\end{equation}
with elastic moment \textcolor{black}{$\lambda_g=N(\mu_q+\mu_{\bar c})=N(\mu_q-\mu_c)$} in the naive-quark model with no spin flip. \textcolor{black}{The same sum rule and the heavy-quark spin symmetry that fix $\lambda$ also fix the elastic vector moment
\cite{Wise1992,AmundsonBoydJenkinsLukeManoharPolitzerWise1992,ChoWise1994,CasalbuoniDeandreaDiBartolomeoGattoFeruglioNardulli1997}.}
The numerical values are
\begin{eqnarray}
\lambda_g(\Dbs{}^0)=N(\mu_u-\mu_c)=+1.13\GeV^{-1},
\nonumber\\
\lambda_g(D^{*-})=N(\mu_d-\mu_c)=-1.08\GeV^{-1}.
\label{eq:lambda-g}
\end{eqnarray}
The $J/\psi$ has no magnetic moment by $C$ parity, so the elastic $J/\psi$ loop vanishes.

\paragraph{Baryon side.} The three magnetic forms are
\begin{align}
-i\,t_{B\to B\gamma} &=
-\,i\,e\,\tilde\mu_B\,\boldsymbol\sigma\!\cdot\!(\bK\!\times\!\beps_\gamma^{*}),
\label{eq:emB}\\
-i\,t_{B^{*}\to B\gamma} &=
-\,i\,e\,\tilde\mu_3\,\bm S_B^{\dagger}\!\cdot\!(\bK\!\times\!\beps_\gamma^{*}),
\label{eq:emBs}\\
-i\,t_{B^{*}\to B^{*}\gamma} &=
-\,i\,e\,\tilde\mu_{B^{*}}\,\bm J\!\cdot\!(\bK\!\times\!\beps_\gamma^{*}),
\label{eq:emBss}
\end{align}
with $\bm S_B$ the $\half\!\to\!\tfrac32$ transition operator of
Eq.~\eqref{eq:SSdag}, $\bm J$ the spin-$\tfrac32$ operator on the $\Scs$ states, and $\tilde\mu\equiv\mu/2m_N$ throughout. Each is a pure $M1$ vertex of the same $(\bK\!\times\!\beps^{*})$ form as the meson vertices, so every triangle ultimately reduces to the universal structure $\mathcal{T}\equiv\chi_f^{\dagger}\,\bm S\!\cdot\!(\bK\!\times\!\beps^{*})\,\chi_i$ of Eq.~\eqref{eq:Mstructure}.

In the naive-quark model the relevant magnetic moments read \textcolor{black}{\cite{WangChenMaLiuZhu2016,LiLiuSunChen2021,Ozdem2021}}
\begin{eqnarray}
\mu(\Sc^{+}\!\to\!\Lc^{+})&=&-\,\frac{\mu_u-\mu_d}{\sqrt3} =-1.63\,\mu_N,
\nonumber\\
\mu({\Scs}^{+}\!\to\!\Lc^{+})&=&\sqrt2\,\mu(\Sc^{+}\!\to\!\Lc^{+})=-2.31\,\mu_N,
\nonumber\\
\mu({\Scs}^{++}\!\to\!\Sc^{++})&=&\frac{2(\mu_u-\mu_c)}{\sqrt3}=+1.67\,\mu_N,
\nonumber\\
\mu({\Scs}^{+}\!\to\!\Sc^{+})&=&\frac{\mu_u+\mu_d-2\mu_c}{\sqrt3}=+0.04\,\mu_N .
\label{eq:mu-trans}
\end{eqnarray}
For the baryon transition magnetic moments, they are
\begin{eqnarray}
\mu_{\Sc^{++}}&=&\frac{4\mu_u-\mu_c}{3}=+2.34\,\mu_N,
\nonumber\\
\mu_{\Sc^{+}}&=&\frac{2\mu_u+2\mu_d-\mu_c}{3}=+0.45\,\mu_N,
\nonumber\\
\mu_{\Lc^{+}}&=&\mu_c=+0.40\,\mu_N,
\nonumber\\
\mu_p&=&+2.79\,\mu_N,
\label{eq:mu-elas-half}
\\
\mu_{{\Scs}^{++}}&=&2\mu_u+\mu_c=+4.11\,\mu_N,
\nonumber\\
\mu_{{\Scs}^{+}}&=&\mu_u+\mu_d+\mu_c=+1.28\,\mu_N,
\label{eq:mu-elas-three-half}
\end{eqnarray}
for the elastic spin-$\half$ and spin-$\tfrac32$ magnetic moments, respectively. The sign of the baryon transition magnetic moments relative to the residue basis is not fixed by the quark model and is the main source of uncertainty in the width.

%---------------------------------------------------------------------
\subsection{Appearance of magnetic vertices in loop diagrams}
\label{app:gauge-conv}

The magnetic vertices of Sec.~\ref{app:em-vertices} are not the only structures that can sit on an internal line. On any elastic radiation ($V\!\to\! V\gamma$ or $B\!\to\! B\gamma$) the photon can also couple convectively (minimally) through the electric charge, in addition to the magnetic structures above. On a transition radiation ($\Dbs\!\to\!\Db\gamma$, $J/\psi\!\to\!\eta_c\gamma$, $\Scs\!\to\!\Sc\gamma$, etc.)
the hadron itself changes and there is no convective coupling, only the magnetic transition vertex of Sec.~\ref{app:em-vertices}. The elastic loops are (g), (g$'$), (g$''$) on the meson side and (b), (k), (f), (l) on the baryon side. In their charged channels they carry both pieces. We show here that the convective piece drops out and only the magnetic piece survives.

Take loop (g) as the example, $\Dbs\Sc\!\to\![\Dbs\,\mathrm{el.}]\!\to\!\Dbs\Sc$.
The full $\Dbs\!\to\!\Dbs\gamma$ vertex on the charged meson line is the sum of the magnetic part of Eq.~\eqref{eq:VVgamma-mag} and the convective part
\begin{equation}
-i\,t_{V\to V\gamma}^{(\mathrm{conv})}
=-\,i\,e\,Q_V\,(p+p')\!\cdot\!\epsilon_{\gamma}^{*}\,(\epsilon_V^{*}\!\cdot\!\epsilon_V^\prime)\,,
\label{eq:conv}
\end{equation}
where $p$ is the meson momentum before and $p'=p-K$ after the photon, and
$Q_V$ is the charge in units of $e$. The analogous decomposition for elastic
$B\!\to\! B\gamma$ on a charged baryon line is the Dirac-plus-Pauli structure,
the Pauli part being Eq.~\eqref{eq:emB} and the Dirac part producing the same
$(p+p')^{\mu}$ convective vertex on the baryon. The argument below treats
both cases at once and shows that the $(p+p')^{\mu}$ piece never reaches the
amplitude.

\paragraph{Longitudinal part and the Ward identity.}
Contracting Eq.~\eqref{eq:conv} with $K_\mu$ uses
\begin{equation}
K\!\cdot\!\,(p+p')
=p^{2}-p'^{2}
=D^{-1}(p)-D^{-1}(p'),
\label{eq:ward}
\end{equation}
the standard Ward identity for the propagator inverse $D^{-1}(p)=p^2-m^2+i\eps$
\textcolor{black}{\cite{Haberzettl:1997jg,Davidson:2001rk,Haberzettl:2021vmd}}.
Inside the loop, this difference cancels the adjacent propagator at one of the
two ends of the radiating line. The two leftover loop integrals are the
two-point functions of the strong vertices at the two ends and they cancel
between each other for an elastic line. The convective (minimal) coupling therefore does not have a transverse $M1$ piece of its own.

\paragraph{Transverse part and Coulomb-gauge kinematics.}
The transverse part of $(p+p')^{\mu}$ drops separately in our kinematics. We
work in the $\Pcp$ rest frame, $P=(M_i,\mathbf 0)$, and the Coulomb gauge
$\eps_\gamma^{0}=0$, $\bm K \cdot\,\bm \eps_\gamma=0$. The time component
$(p+p')^{0}$ multiplies $\eps_\gamma^{0}=0$ and so it vanishes. The space component
is $(\bm p+\bm p')\cdot \beps_{\gamma}$. The two strong vertices and the spectator
line do not depend on the loop momentum, so the loop integral of $(\bm p+ \bm p')$
is a three-vector built only from the external momenta. In the $\Pcp$ rest
frame $\bm P=\mathbf 0$, so the only available three-vector is $\bm K$, and
\begin{equation}
\int\!\frac{d^{4}q}{(2\pi)^{4}}\,(\bm p+ \bm p')\,f(q)\;\propto\;\bm K\,,
\qquad \bm K\!\cdot\!\beps_\gamma=0.
\label{eq:convzero}
\end{equation}
The convective coupling gives zero on both the meson line and the baryon line.

\paragraph{Magnetic term dominance.}
Only the magnetic coupling of Eqs.~\eqref{eq:VVgamma-mag} and
\eqref{eq:emB} \textcolor{black}{to} \eqref{eq:emBss} remains. Being proportional to
$\bK\!\times\!\beps_\gamma^{*}$, it is transverse on its own,
$\bm K\!\cdot\!(\bK\!\times\!\beps_\gamma^{*})=0$, so each triangle is gauge
invariant separately and the result is the same in any gauge. The elastic
loops then keep only $\lambda_g$ on the meson side and $\tilde\mu_B$ or
$\tilde\mu_{B^{*}}$ on the baryon side, the transition loops keep $\lambda$,
$\lambda_\psi$, $\tilde\mu_3$, $\tilde\mu_{\Sc\Lc}$ and
$\tilde\mu_{\Scs\Lc}$, and the sum of the nineteen loops is a clean $M1$ mode.

%---------------------------------------------------------------------
\subsection{Spin factors $\mathfrak{s}_X$ from the baryon line}
\label{app:gauge-baryon}

With every photon vertex magnetic, the spin algebra of each triangle reduces to a single scalar $\mathfrak{s}_X$ multiplying $\mathcal T$. We treat the cases in which $\mathfrak{s}_X$ is determined by direct evaluation, and mark with a dagger ($^{\dagger}$) the six loops that carry both a $\Scs$ and a
$\Dbs$ in the triangle. The latter involve a $\tfrac32\otimes 1$ recoupling whose leading-order value is fixed by analogy in App.~\ref{app:spin}.

\paragraph{$B\to B\gamma$, spin-$\tfrac12$ elastic.}
The spectator $\Dbs$ ties the initial $\bm{S}\!\cdot\!\beps_V$ vertex of Eq.~\eqref{eq:v32V} to the final $\bm{\sigma}\!\cdot\!\beps_V$ vertex of Eq.~\eqref{eq:v12V} through the polarization sum $\sum\eps_V^{i}\eps_V^{*j}=\delta^{ij}$, and the elastic
$\bm{\sigma}\!\cdot\!(\bK\!\times\!\beps^{*})$ of Eq.~\eqref{eq:emB} sits on the baryon between them. The baryon chain is $\chi_f^{\dagger}\,\bm{\sigma}\!\cdot\!\big[\boldsymbol\sigma\!\cdot\!(\bK\!\times\!\beps^{*})\big]\!\bm{S}\,\chi_i$,
and with the identity $\bm{\sigma}\!\cdot\!(\boldsymbol\sigma\!\cdot\!\bm A)\!\bm S
=2\bm S\!\cdot\!\bm A$ of App.~\ref{app:spin} it returns $+2\,\mathcal T$. This is the $+2$ entry of loops (b), (e), (e$'$), (f) and (k).

\paragraph{$B^{*}\!\to\! B\gamma$, spin-$\tfrac32\to\tfrac12$ transition.}
With a pseudoscalar spectator, loops (d) and (m), the initial vertex Eq.~\eqref{eq:v32P} is the identity on the $\tfrac32$ space and the final vertex Eq.~\eqref{eq:v12P} is the scalar. The baryon chain reduces to $\chi_f^{\dagger}\,\bm S_B^{\dagger}\!\cdot\!(\bK\!\times\!\beps^{*})\,\chi_i$, and since $\bm S_B$ satisfies the same algebra as $\bm S$ in Eq.~\eqref{eq:SSdag}, this is exactly $+\mathcal T$, the entry $+1$. With a vector spectator, loops (h) and (i), the spectator $\Dbs$ ties the $\tfrac32$
vertex $\bm S^{\dagger}\!\cdot\!\sigma_j \bm S$ at the initial end to the $\boldsymbol\sigma$ vertex at the final end while $\bm S_B^{\dagger}$ acts on the baryon between a $\Scs$ and a $\Sc$. This is a $\tfrac32\otimes 1$ recoupling whose leading-order value is $+\mathcal T$, the entry $+1^{\dagger}$.

\paragraph{$B^{*}\!\to\! B^{*}\gamma$, spin-$\tfrac32$ elastic.}
The single loop of this kind is (l),
$\Dbs\Scs\!\to\![\Scs\,\mathrm{el.}]\!\to\!\Dbs\Scs$. The vertex uses the
spin-$\tfrac32$ operator $\bm J$ of Eq.~\eqref{eq:emBss} with matrix element
$\langle m'|\bm J|m\rangle$ between $\Scs$ states, the spin-$\tfrac32$
counterpart of the vector-meson relation
$\langle\beps_V'|\bm S^{(1)}|\beps_V\rangle=-i(\beps_V^{\prime*}\!\times\!\beps_V)$
used for the elastic meson loops in App.~\ref{app:spin}. The spectator $\Dbs$
ties the two $\tfrac32$ vertices and a second $\Scs$ enters at the final end,
so the reduction is once more a $\tfrac32\otimes 1$ recoupling. At leading
order it gives $+\mathcal T$, the entry $+1^{\dagger}$. There is no companion
loop with a pseudoscalar spectator, because $\Db\Scs$ couples to $\tfrac32$
and not to the $\half$ of the $\Pc(4312)$.

\paragraph{\textcolor{black}{Meson-baryon} mirror.}
$B^{*}\!\to\! B\gamma$ and $B^{*}\!\to\! B^{*}\gamma$ are the baryon mirror of
$V\!\to\! P\gamma$ and $V\!\to\! V\gamma$. All four reduce to the single
structure $\mathcal T$, so they are not independent tensors in the amplitude.
They differ in two places. The transition moment $\tilde\mu_3$ and the
elastic moment $\tilde\mu_{B^{*}}$ are different combinations of the same
constituent moments $\mu_u,\mu_d,\mu_c$ of Eqs.~\eqref{eq:mu-trans} \textcolor{black}{to}
\eqref{eq:mu-elas-three-half}, fixed by the naive-quark model and tied by
heavy-quark spin symmetry, in the same way that $\lambda$ and $\lambda_g$ are
on the meson side. The spin factor is then set by the spin chain: $+1$ for
the transition with a pseudoscalar spectator and the leading-order $+1$ for
the vector-spectator and elastic cases. The six loops marked with a dagger
keep that label because their $\tfrac32\otimes 1$ recoupling is taken at
leading order. A simultaneous change of all six moves the width by $2.7\%$
(App.~\ref{app:spin}), so these estimates do not affect the result.

\section{Loop master formula and analytic reduction of the Feynman-parameter integral}
\label{app:integral}
\label{app:yint}

This appendix gives the loop-integration formula used in the numerical
evaluation. We also give the analytic form obtained after carrying out the
inner Feynman-parameter integral. The purpose is to make the connection between
the four-dimensional triangle integral and the one-dimensional numerical
quadrature used in the main text explicit.

For a triangle diagram with three propagators, we combine the denominators as
\begin{equation}
\frac{1}{D_1D_2D_3}
=
2\int_0^1 dx\int_0^{1-x}dy\,
\frac{1}{\left[x_1D_1+xD_2+yD_3\right]^3},
\label{eq:feynman_app}
\end{equation}
where $x_1=1-x-y$. With the momentum routing used in the main text, the Feynman denominator can be
written as
\begin{equation}
x_1D_1+xD_2+yD_3
=
(q-\bar q)^2-\Delta_{X,\alpha}(x,y)+i\epsilon ,
\label{eq:complete_square_app}
\end{equation}
where
\begin{equation}
\bar q=(1-y)P-xK,
\qquad
\Delta_{X,\alpha}(x,y)=-s_{X,\alpha}(x,y).
\label{eq:qbar_delta_app}
\end{equation}
The function $s_{X,\alpha}$ is the quadratic form given in
Eq.~\eqref{eq:smaster}. After shifting $q\to \ell+\bar q$ and performing the
Wick rotation, the scalar master integral is
\begin{eqnarray}
&&i\int\frac{d^4\ell}{(2\pi)^4}
\frac{1}{\left[\ell^2-\Delta_{X,\alpha}(x,y)+i\epsilon\right]^3}
\nonumber\\
&&\qquad = \frac{1}{32\pi^2}
\frac{1}{\Delta_{X,\alpha}(x,y)}=
-\frac{1}{32\pi^2}
\frac{1}{s_{X,\alpha}(x,y)} .
\label{eq:master_app}
\end{eqnarray}
Together with the overall Feynman-parameter factor in
Eq.~\eqref{eq:feynman_app}, this gives the common factor
$1/(16\pi^2)$ in the amplitudes.

The numerator determines the Feynman-parameter weight. For a meson-line photon,
the magnetic $V\to P\gamma$ numerator is linear in the radiating-meson momentum.
Using
\begin{equation}
P-q=yP+xK-\ell ,
\label{eq:pminusq_app}
\end{equation}
the term odd in $\ell$ integrates to zero. The term proportional to $K$ vanishes
after contraction with
$\varepsilon_{\mu\nu\alpha\beta}K^\beta$. Hence only the $yP$ term survives.
This gives
\begin{equation}
w_M(y)=y .
\label{eq:wM_app}
\end{equation}
For a baryon-line photon, the magnetic vertex is independent of the loop
momentum at this order. Therefore
\begin{equation}
w_B(y)=1 .
\label{eq:wB_app}
\end{equation}
Thus the two kinds of loop integrals needed in this work are
\begin{equation}
\mathcal I_{X,\alpha}[w_\kappa]
=
\int_0^1 dx\int_0^{1-x}dy\,
\frac{w_\kappa(y)}{s_{X,\alpha}(x,y)},
\quad
\kappa=M,B .
\label{eq:Iwkappa_app}
\end{equation}

At fixed $x$, the denominator is a quadratic polynomial in $y$,
\begin{equation}
s_{X,\alpha}(x,y)
=
a\,y^2+b_{X,\alpha}(x)\,y+c_{X,\alpha}(x),
\quad
a=-M_i^2 .
\label{eq:quadratic_app}
\end{equation}
The two roots are
\begin{eqnarray}
r_{1,2}^{X,\alpha}(x)
&=&
\frac{
-b_{X,\alpha}(x)
\pm
\sqrt{b_{X,\alpha}^2(x)-4a\,c_{X,\alpha}(x)}
}{2a}\,,
\nonumber\\
s_{X,\alpha}
&=&
a\left(y-r_1^{X,\alpha}\right)
 \left(y-r_2^{X,\alpha}\right).
\label{eq:roots_app}
\end{eqnarray}
The square-root branch is fixed by the $+i\epsilon$ prescription in
$c_{X,\alpha}$. For a closed channel the radicand is real, so the roots are real
or form a complex-conjugate pair. For an open channel the same prescription
moves the pole away from the integration contour and produces the absorptive
part of the loop.

Define
\begin{equation}
\mathcal L_k^{X,\alpha}(x)
=
\ln
\frac{
1-x-r_k^{X,\alpha}(x)
}{
-r_k^{X,\alpha}(x)
},
\qquad
k=1,2 .
\label{eq:Lk_app}
\end{equation}
Then the inner $y$ integration can be done analytically. For the meson-line
weight one obtains
\begin{eqnarray}
I_M^{X,\alpha}(x)
&\equiv&
\int_0^{1-x}dy\,
\frac{y}{s_{X,\alpha}(x,y)}
\nonumber\\
&=&
\frac{
r_1^{X,\alpha}(x)\mathcal L_1^{X,\alpha}(x)
-
r_2^{X,\alpha}(x)\mathcal L_2^{X,\alpha}(x)
}{
a\left[
r_1^{X,\alpha}(x)-r_2^{X,\alpha}(x)
\right]
}.
\label{eq:IM_app}
\end{eqnarray}
For the baryon-line weight one obtains
\begin{eqnarray}
I_B^{X,\alpha}(x)
&\equiv&
\int_0^{1-x}dy\,
\frac{1}{s_{X,\alpha}(x,y)}
\nonumber\\
&=&
\frac{
\mathcal L_1^{X,\alpha}(x)
-
\mathcal L_2^{X,\alpha}(x)
}{
a\left[
r_1^{X,\alpha}(x)-r_2^{X,\alpha}(x)
\right]
}.
\label{eq:IB_app}
\end{eqnarray}
These expressions reduce the triangle amplitude to a single numerical
quadrature over $x$. In the notation of the main text,
\begin{equation}
A_{(X)}
=
\mathfrak s_X\,
(g_ig_f)_X\,
\frac{(M_i)^{\delta_{\kappa M}}}{16\pi^2}
\sum_\alpha
\mathcal C_\alpha^2\,
\Xi_{X,\alpha}^{(\kappa)}
\int_0^1 dx\,
I_\kappa^{X,\alpha}(x)\,,
\label{eq:Afinal_app}
\end{equation}
with $\kappa=M,B$. The factor $M_i$ appears only in the meson-line
case because the surviving part of the covariant $V\to P\gamma$ numerator is
proportional to $P^\nu$ in the initial-rest frame. The baryon-line magnetic
vertices already contain the external magnetic field and do not give this
additional factor.

For a numerical check, consider the neutral part of loop (a), with
\begin{eqnarray}
(M_1,M_2,M_s)
&=&
(2.00685,\,1.86484,\,2.45290)\ {\rm GeV},
\nonumber\\
E_\gamma &=& 0.14303\ {\rm GeV}.
\end{eqnarray}
The quadratic coefficients are
\begin{eqnarray}
&&a=-19.8675,\qquad
\textcolor{black}{b_a^{(n)}(x)=17.8782-1.2751\,x,}
\nonumber\\
&&\textcolor{black}{c_a^{(n)}(x) = -4.0275+0.5498\,x ,}
\label{eq:numerical_coeff_app}
\end{eqnarray}
in units of ${\rm GeV}^2$. At $(x,y)=(0.2,0.3)$,
\begin{equation}
s_a^{(n)}=-0.41859\ {\rm GeV}^2,
\qquad
\frac{y}{s_a^{(n)}}=-0.7167\ {\rm GeV}^{-2}.
\end{equation}
At fixed $x=0.2$, Eq.~\eqref{eq:roots_app} gives
\begin{equation}
r_{1,2}=0.4435\mp0.0217\,i .
\end{equation}
Using Eq.~\eqref{eq:IM_app} and then integrating over $x$ gives
\begin{equation}
\int_0^1 dx\,I_M^{a,n}(x)
=
-1.7048\ {\rm GeV}^{-2}.
\label{eq:numerical_check_app}
\end{equation}
Multiplying this result by the neutral prefactor and adding the charged
contribution reproduces the loop amplitude quoted in the main text.

\section{Width formula and open-channel normalization}
\label{app:width}

Squaring Eq.~\eqref{eq:Mstructure}, summing photon polarizations and averaging
over the four initial spin states gives Eq.~\eqref{eq:width},
$\Gamma=\tfrac{4\alpha}{3}\tfrac{M_f}{M_i}E_\gamma^3|\tilde A|^2$. \textcolor{black}{The
factor $M_f/M_i$ and the factor four come from the non-relativistic baryon
normalization $\bar u u=2M_B$ on the two baryon legs and the average over the four
$\tfrac32^{-}$ spin states. The meson coupling $\lambda$ in $\tilde A$ is fixed
from the free decay $D^{*}\to D\gamma$ of Eq.~\eqref{eq:lamdef}.} For a resonance $R$ coupling
to a meson-baryon channel with dimensionless residue $g$,
\begin{equation}
\Gamma_i=\frac{M_B}{2\pi M_R}\,|g_i|^{2}\,k_i,
\qquad k_i=\frac{\lambda^{1/2}(M_R^2,m_M^2,M_B^2)}{2M_R}.
\label{eq:openwidth}
\end{equation}
This fixes the $J/\psi p$ branchings used in Sec.~\ref{sec:lambdab}. For the
$P_c(4457)$, $\Gamma_{J/\psi p}=2.48\MeV$ of a model total of $3.0\MeV$, so
$\mathcal B=0.829$. \textcolor{black}{For the $P_c(4312)$, $\Gamma_{J/\psi p}=4.84\MeV$
and $\Gamma_{\eta_c N}=12.26\MeV$. The pole total $\Gamma_{P_c(4312)}=2\,{\rm Im}\sqrt{s_0}=15.2\MeV$
includes every channel. These two partials alone exceed it, and the open
$\bar D\Lambda_c$ and $\bar D^{*}\Lambda_c$ channels add more, so the on-shell
two-body formula overestimates the partials. We take
$\mathcal B(P_c(4312)\to J/\psi p)\simeq0.28$ as an upper estimate.} The same check fixes the $1/\sqrt3$ of Eq.~\eqref{eq:v12V} and the
$1/\sqrt2$ of Eq.~\eqref{eq:v12Vs}. The inputs used are $\alpha=1/137.036$,
$m_N=938.27\MeV$, $\mathcal B(\Lambda_b\to J/\psi pK)=3.2\times10^{-4}$
\cite{PDG2024}, the LHCb fit fractions $R_{P_c(4457)}=0.53\%$ and
$R_{P_c(4312)}=0.30\%$, and $\Gamma_{P_c(4457)}^{\rm exp}=6.4\pm2.0\MeV$
\cite{LHCb2019}, the quark magnetic moments $\mu_u=1.852$, $\mu_d=-0.972$,
$\mu_c=0.404\,\mu_N$ \cite{PDG2024}, and the calibration
$N=0.783\GeV^{-1}/\mu_N$.

\bibliographystyle{apsrev4-2}
\bibliography{Ref_Pc_radiative_PRD}

\end{document}